\documentclass[a4paper,11pt]{article}
\pdfoutput=1 
\usepackage{jinstpub}
\usepackage{caption}
\usepackage[labelformat=simple]{subcaption}

\usepackage{indentfirst}
\usepackage[version-1-compatibility,load=addn, binary-units=true]{siunitx}
\DeclareSIUnit[number-unit-product = ]\percent{\char`\%}
\usepackage[acronym,nomain]{glossaries}
\usepackage[symbol]{footmisc}
\renewcommand{\thefootnote}{\fnsymbol{footnote}}
\usepackage{lineno}
\usepackage[sort&compress]{natbib}
\newacronym{afe}{AFE}{analogue front-end}
\newacronym{asic}{ASIC}{application-specific integrated circuit}
\newacronym{bx}{BX}{bunch crossing}
\newacronym{be}{BE}{back-end}
\newacronym{cdr}{CDR}{clock data recovery}
\newacronym{cml}{CML}{current mode logic}
\newacronym{cms}{CMS}{Compact Muon Solenoid}
\newacronym{csa}{CSA}{charge sensitive amplifier}
\newacronym{da}{DA}{differential amplifier}
\newacronym{dac}{DAC}{digital-to-analogue converter}
\newacronym{daq}{DAQ}{data acquisition}
\newacronym{diff}{DIFF}{differential}
\newacronym{enc}{ENC}{equivalent noise charge}
\newacronym{fe}{FE}{front-end}
\newacronym{hdi}{HDI}{high density interconnect}
\newacronym{hllhc}{HL-LHC}{High Luminosity LHC}
\newacronym{it}{IT}{Inner Tracker}
\newacronym{ip}{IP}{intellectual property}
\newacronym{lcc}{LCC}{leakage current compensation}
\newacronym{led}{LED}{light-emitting diodes}
\newacronym{lhc}{LHC}{Large Hadron Collider}
\newacronym{lin}{LIN}{linear}
\newacronym{ls3}{LS3}{Long Shutdown 3}
\newacronym{mip}{MIP}{minimum ionizing particle}
\newacronym{mpv}{MPV}{most probable value}
\newacronym{ot}{OT}{Outer Tracker}
\newacronym{pa}{PA}{preamplifier}
\newacronym{pll}{PLL}{phase locked loop}
\newacronym{rms}{RMS}{root-mean-square}
\newacronym{sync}{SYNC}{synchronous}
\newacronym{shldo}{ShLDO}{shunt low-dropout}
\newacronym{tbpx}{TBPX}{Tracker Barrel Pixel detector}
\newacronym{tepx}{TEPX}{Tracker Endcap Pixel detector}
\newacronym{tfpx}{TFPX}{Tracker Forward Pixel detector}
\newacronym{tia}{TIA}{transimpedance amplifier}
\newacronym{tid}{TID}{total ionizing dose}
\newacronym{tot}{TOT}{time-over-threshold}
\newacronym{tsmc}{TSMC}{Taiwan Semiconductor Manufacturing Company}

\newcommand\blfootnote[1]{%
  \begingroup
  \renewcommand\thefootnote{}\footnote{#1}%
  \addtocounter{footnote}{-1}%
  \endgroup
}

\title{Comparative evaluation of analogue front-end designs for the CMS Inner Tracker at the High Luminosity LHC}

\emailAdd{natalia.emriskova@cern.ch}

\abstract{
The CMS Inner Tracker, made of silicon pixel modules, will be entirely replaced prior to the start of the High Luminosity LHC period. One of the crucial components of the new Inner Tracker system is the readout chip, being developed by the RD53 Collaboration, and in particular its analogue front-end, which receives the signal from the sensor and digitizes it. Three different analogue front-ends (Synchronous, Linear, and Differential) were designed and implemented in the RD53A demonstrator chip. A dedicated evaluation program was carried out to select the most suitable design to build a radiation tolerant pixel detector able to sustain high particle rates with high efficiency and a small fraction of spurious pixel hits. The test results showed that all three analogue front-ends presented strong points, but also limitations. The Differential front-end demonstrated very low noise, but the threshold tuning became problematic after irradiation. Moreover, a saturation in the preamplifier feedback loop affected the return of the signal to baseline and thus increased the dead time. The Synchronous front-end showed very good timing performance, but also higher noise. For the Linear front-end all of the parameters were within specification, although this design had the largest time walk. This limitation was addressed and mitigated in an improved design.
The analysis of the advantages and disadvantages of the three front-ends in the context of the CMS Inner Tracker operation requirements led to the selection of the improved design Linear front-end for integration in the final CMS readout chip.}


\collaboration{\includegraphics[height=20mm]{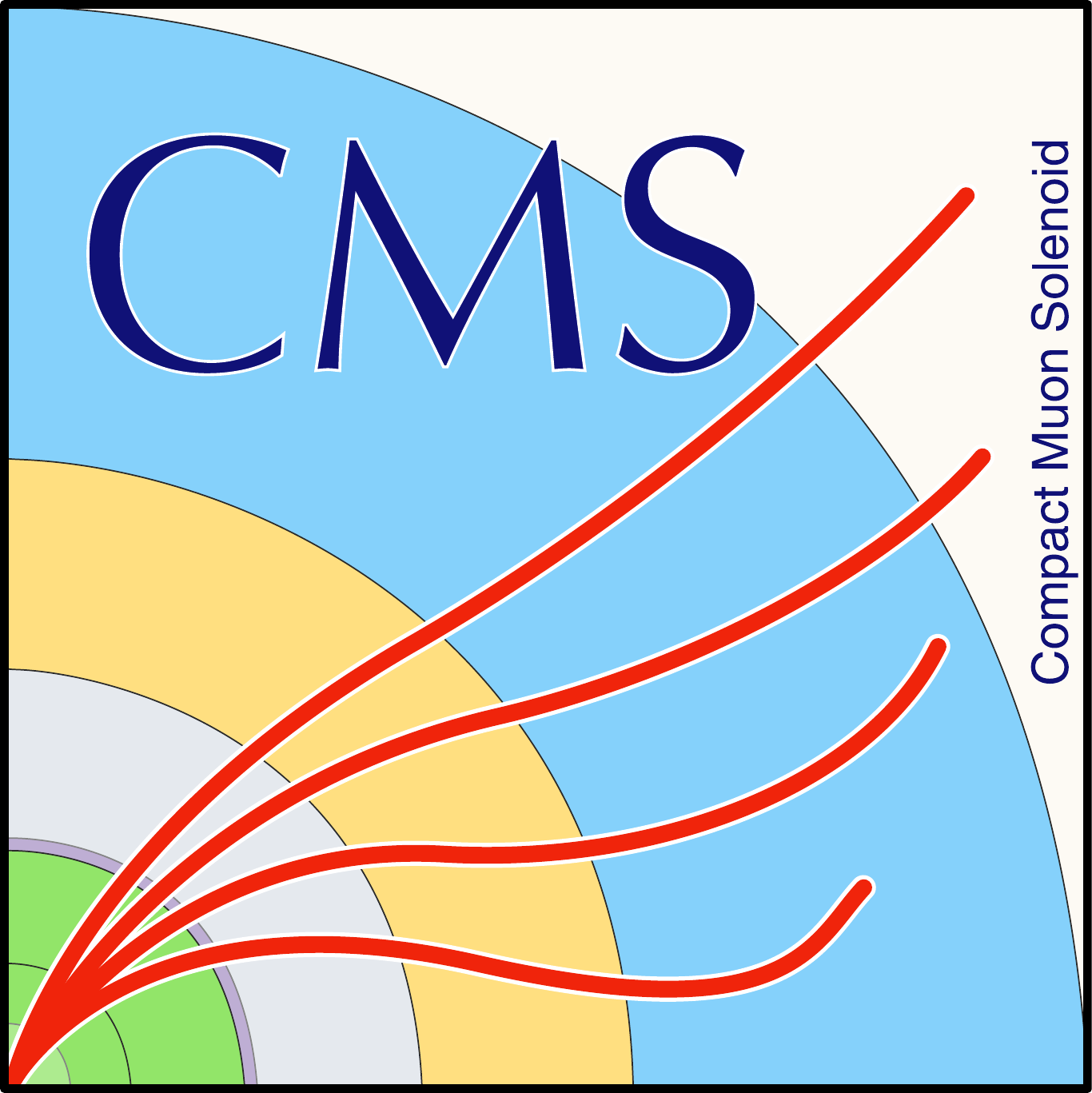}\\[12pt]
The Tracker Group of the CMS Collaboration\blfootnote{Corresponding author: Natalia Emriskova}}

\glsdisablehyper

\begin{document}
    \maketitle
    \flushbottom
\section{CMS pixel detector upgrade for the High Luminosity LHC}
\label{sec:intro}

The High Luminosity upgrade~\citep{hllhc} of the CERN \gls{lhc}~\citep{lhc} will boost its potential for physics discoveries, but also impose extreme operating conditions for the experiments. Along with the accelerator, the \gls{cms}~\citep{cms_exp} detector will be substantially upgraded during \acrlong{ls3}, starting in \num{2025}~\citep{hllhc_web}. This upgrade is referred to as the \gls{cms} Phase-2 Upgrade~\citep{cms_p2}.
The silicon tracking system, located at the heart of \gls{cms}, detects trajectories of charged particles. It will be entirely replaced during the \acrlong{ls3} because of the accumulated radiation damage and to take advantage of the increased luminosity. The goal of the upgrade is to maintain or improve the tracking and vertex reconstruction performance of the detector in the harsh environment of the \gls{hllhc}. The \gls{cms} Phase-2 tracker will consist of the \acrlong{ot}, made of silicon modules with strip and macro-pixel sensors, and the \gls{it}, based on silicon pixel modules~\citep{p2_tdr}.

\begin{figure}[t]
    \centering
    \includegraphics[width=0.8\textwidth]{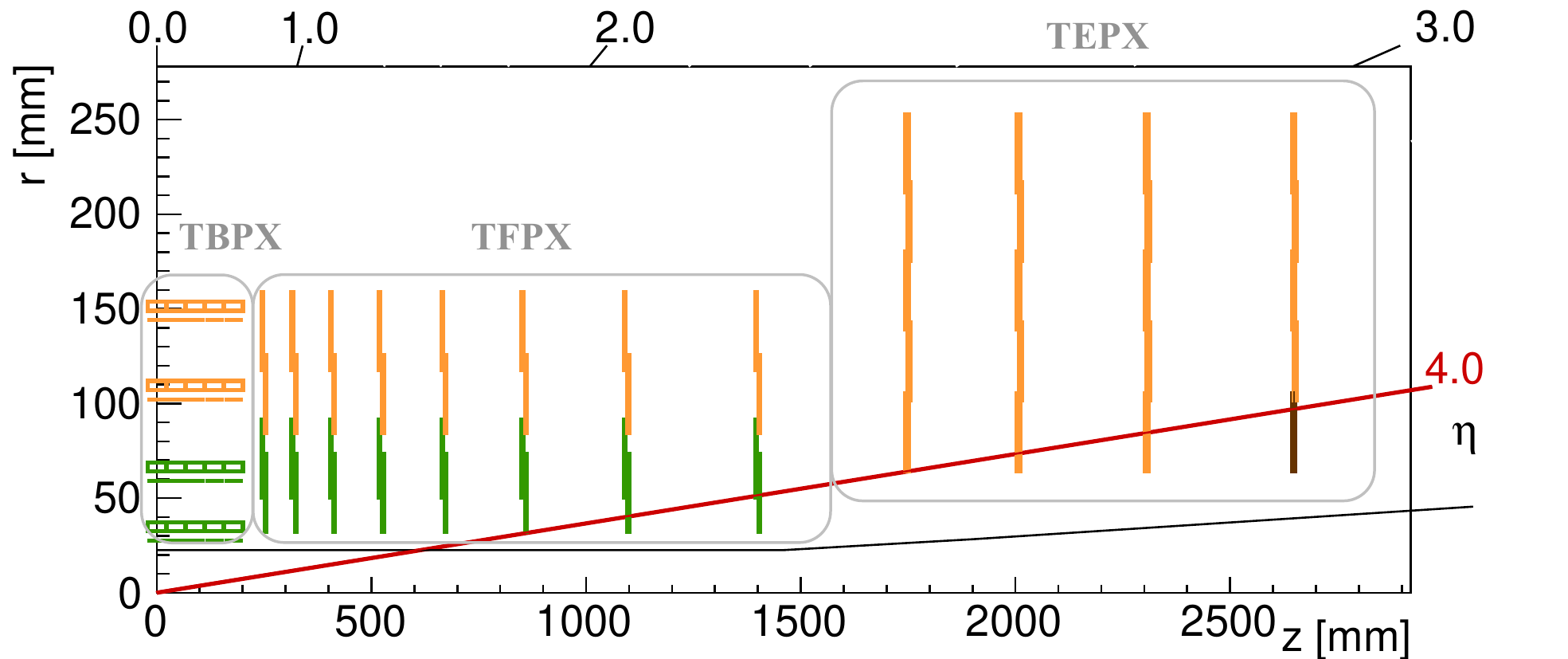}
    \caption{Layout of one quarter of the Phase-2 Inner Tracker in the $r$-$z$ view. Green lines correspond to pixel modules with two readout chips and orange lines represent modules with four chips. The modules shown in brown correspond to the innermost ring of the last \gls{tepx} disc ($z=\SI{2650}{\milli\meter}$), which will be used by the Beam Radiation Instrumentation and Luminosity (BRIL) project~\citep{bril_cdr} for dedicated luminosity and background measurements. The grey line represents the beam pipe envelope.}
    \label{fig:itlayout}
\end{figure}

The high granularity of the \gls{it} offers excellent spatial resolution, which is important for a precise three-dimensional reconstruction of particle trajectories, as well as the identification of primary interaction vertices and secondary decay vertices. 
One quarter of the latest layout of the Phase-2 \gls{it} in the $r$-$z$\footnote{\gls{cms} adopts a right-handed coordinate system with the origin centred at the nominal collision point inside the experiment. The $x$ axis points towards the centre of the \gls{lhc}, the $y$ axis points vertically upwards and the $z$ axis points along the beam direction. The azimuthal angle $\phi$ is measured from the $x$ axis in the $x$-$y$ plane, the radial coordinate in this plane is denoted by $r$ and the polar angle $\theta$ is measured from the $z$ axis. The pseudorapidity $\eta$ is defined as $\eta = -ln\tan (\theta/2)$~\citep{p2_tdr}.} 
view is shown in Figure~\ref{fig:itlayout}. The \gls{it} will consist of a barrel component with four cylindrical layers, referred to as the \gls{tbpx}. Eight smaller double-discs forming the \gls{tfpx} and four larger double-discs forming the \gls{tepx} will be placed in the forward direction on each side. The forward acceptance will be extended up to a pseudorapidity of $|\eta| = 4$~\citep{p2_tdr}, as indicated by the red line in Figure~\ref{fig:itlayout}.

The \gls{it} detector will have an active area of \SI{4.9}{\meter\squared} and it will be composed of \num{3892} pixel modules. Two pixel sizes are currently being considered for the Phase-2 \gls{it}: \SI[product-units = power]{100x25}{\micro\meter} pixels and \SI[product-units = power]{50x50}{\micro\meter} pixels. With a pixel size of \SI{2500}{\micro\meter\squared} there would be about \num{2}~billion readout channels. The detector design strives for a minimal mass of the detector to avoid degradation of the tracking performance due to the interactions of particles with the detector material. Therefore, lightweight mechanical structures made of carbon fiber, two-phase CO\textsubscript{2} cooling~\citep{p2_tdr} and a low voltage powering scheme based on serial powering~\citep{sp_vertex19} will be used. The data will be transmitted through low-mass electrical links and optical fibers~\citep{p2_tdr} to further reduce the detector mass.

\begin{figure}[!ht]
    \centering
    \includegraphics[width=0.6\textwidth]{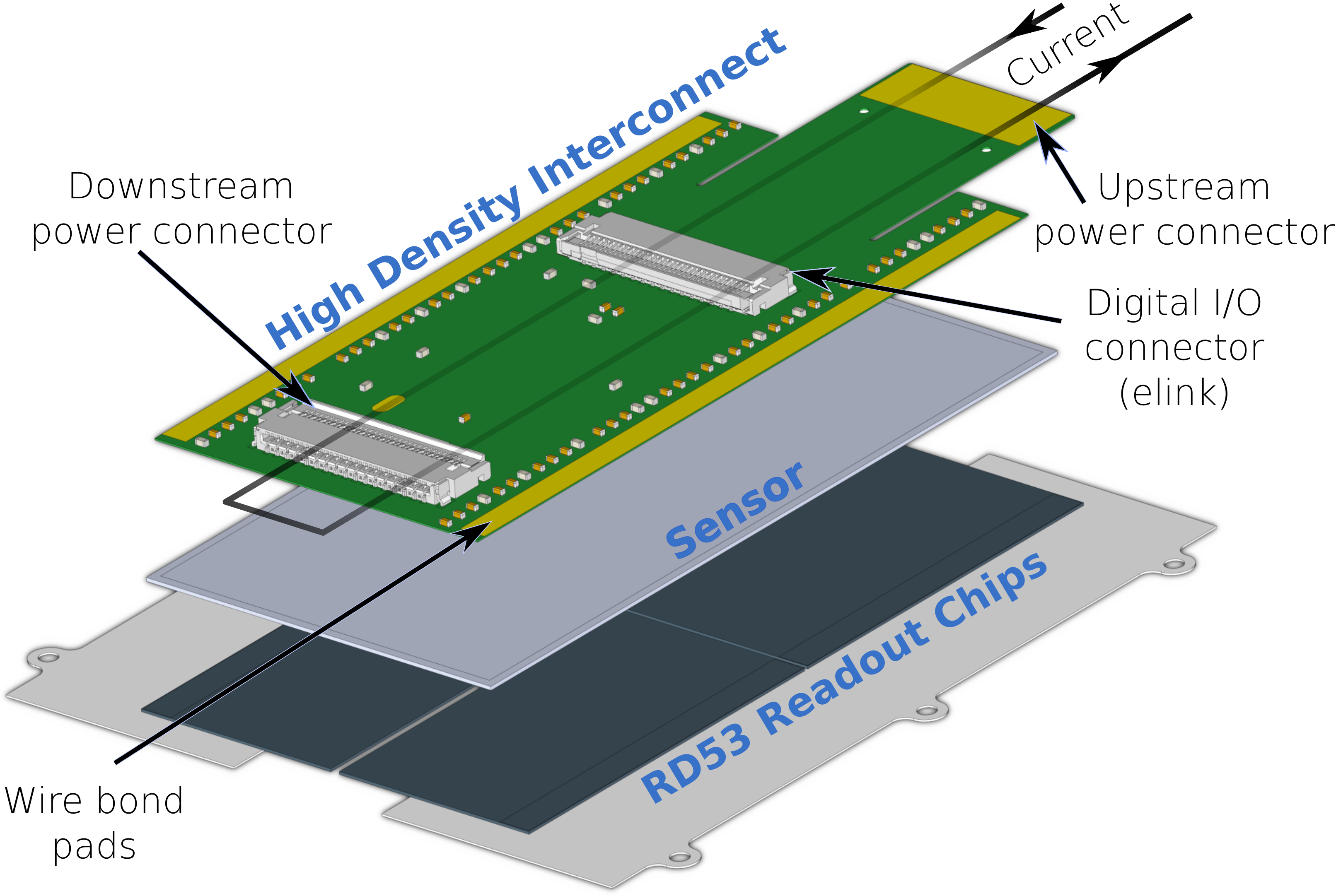}
    \caption{A 3D exploded view of the Phase-2 Inner Tracker pixel module with four readout chips~\citep{bane_module}. From top to bottom the following components can be seen: the high density interconnect, the silicon pixel sensor, the RD53 readout chips, and the rails to mount the module on a support structure and to ensure the electrical isolation of the module.}
    \label{fig:module}
\end{figure}

The main building block of the \gls{it} system is a hybrid pixel module, shown in Figure~\ref{fig:module}. It is composed of a silicon sensor bump-bonded to two or four readout chips. Pixel modules with two readout chips are indicated in green in Figure~\ref{fig:itlayout} and modules with four chips are indicated in orange.
The readout chips of the module are wire-bonded to a flexible printed circuit board with passive components and connectors, called the \acrlong{hdi}, that distributes power (low voltage to power the pixel chips and high voltage to bias the sensor), clock and control signals and collects data from the chips. Signals produced in the sensor are transmitted to the front-end electronics, where they are processed and the data are stored during the planned \SI{12.8}{\micro\second} trigger latency interval. The hit information is sent to the back-end \acrlong{daq} system of the experiment only after receipt of a Level-1 trigger~\citep{L1_trigger} signal. In the innermost layer of the \gls{it}, the hit rate will reach \SI{3.5}{\giga\hertz\per\centi\meter\squared}, while the \gls{cms} Level-1 accept rate will increase from \num{100} to \SI{750}{kHz}~\citep{L1_trigger}.

\section{RD53A analogue front-ends}
\label{sec:pixelchip}

A high-performance radiation tolerant pixel readout chip is essential for good tracking performance of the \gls{it} operating during the \gls{hllhc} era. Such a readout chip is being designed in \acrshort{tsmc}~\cite{tsmc_web} \SI{65}{\nano\meter} CMOS technology by the RD53 Collaboration~\citep{rd53}, a joint effort between the ATLAS and \gls{cms} experiments.
A large-scale demonstrator chip called RD53A~\citep{rd53a_manual} containing design variations was produced. Its purpose is to demonstrate the suitability of the chosen technology for low threshold, low noise, and low power operation at high hit rates, to verify sufficient radiation tolerance~\citep{rd53a_specs}, and to select the most suitable design for the final readout chip.
It is a mixed signal chip, having both analogue and digital circuits. It features custom-designed \acrlong{ip} blocks, such as \acrlong{cdr} and \acrlong{pll} blocks~\citep{rd53a_cdr} for the clock recovery from the command stream running at \SI{160}{\mega b\per\second}; a high speed output transmitter with a \acrlong{cml} cable driver~\citep{rd53a_transmitter} sending data at \SI{1.28}{\giga b\per\second} on up to four output lanes; and a \acrlong{shldo} regulator~\citep{shldo} for serial powering of the pixel modules.
The chip size is \SI[product-units = power]{20.0 x 11.8}{\milli\meter}, which is about half the size of the final chip, as it shares the chip reticle with \gls{cms} \acrlong{ot} chips. The pixel matrix is composed of \num{400 x 192} square pixels with \SI{50}{\micro\metre} pitch. 
All the common analogue and digital circuitry needed to bias, configure, monitor, and read out the chip is placed at the bottom chip periphery~\citep{rd53a_manual}.

\begin{figure}[t]
    \centering
    \includegraphics[width=\textwidth]{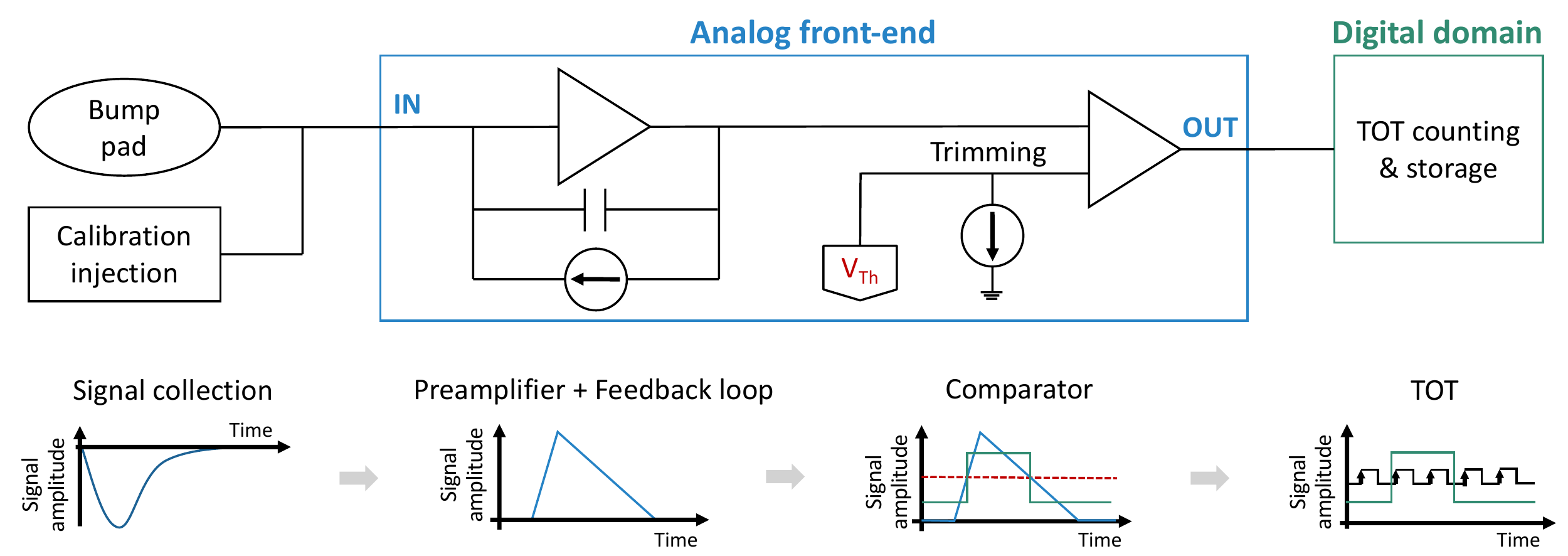}
    \caption{Signal processing steps in different stages of a generic analogue front-end, from signal collection to digitization.}
    \label{fig:afesteps}
\end{figure}

The analogue-to-digital conversion is performed by the \gls{afe}, whose basic structure (shown in Figure~\ref{fig:afesteps}) includes a \gls{csa}, usually referred to as \gls{pa}, a feedback circuit taking care of the signal return to baseline and leakage current compensation, a threshold discriminator, a threshold trimming circuit to address pixel-to-pixel variation of the threshold voltage, and a \gls{tot} counting of the input signal amplitude. In the RD53A chip the \gls{tot}$_{40}$ digitization with 4-bit resolution is done with respect to rising edges of the \SI{40}{\mega\hertz} \gls{lhc} clock\footnote{In the final pixel chip, the counting will be performed on both the rising and the falling edge of the clock, resulting in a finer \gls{tot}$_{80}$ counting at \SI{80}{\mega\hertz} with one \gls{tot}$_{80}$ unit equal to \SI{12.5}{\nano\second} \cite{rd53b_manual}.}. Therefore, one \gls{tot}$_{40}$ unit corresponds to \SI{25}{\nano\second} \cite{rd53a_manual}.
The chip also features a circuit for the generation of internal calibration charge injection signals. The circuit, connected to the input of the \gls{pa}, enables the injection of a well-defined and programmable charge to test the front-end functionalities and calibrate the chip response. Every pixel in the RD53A chip contains the same circuit based on two switches that generate voltage steps fed to an injection capacitor~\citep{rd53a_manual}.

\begin{figure}[ht]
    \captionsetup{justification=centering}
    \begin{minipage}{0.45\textwidth}
        \centering
        \includegraphics[trim=150 0 100 0, clip, height=3.9cm]{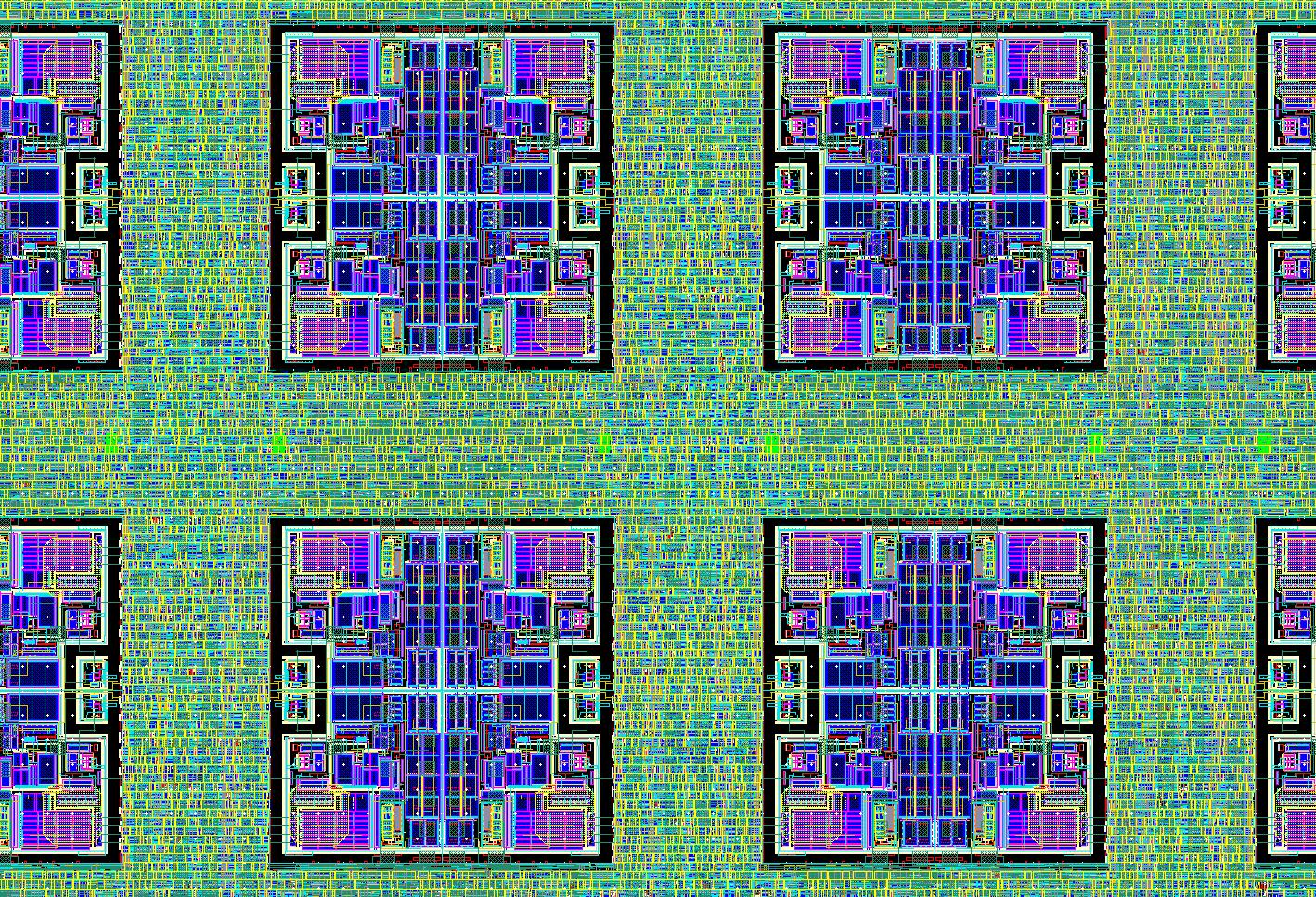}
        \caption{RD53A layout of four analogue islands, i.e.~sixteen pixels, surrounded by the fully synthesized digital “sea”~\citep{rd53a_manual}.}
        \label{fig:ana_island}
    \end{minipage}
    \hfill
    \begin{minipage}{0.5\textwidth}
        \centering
        \includegraphics[width=\textwidth]{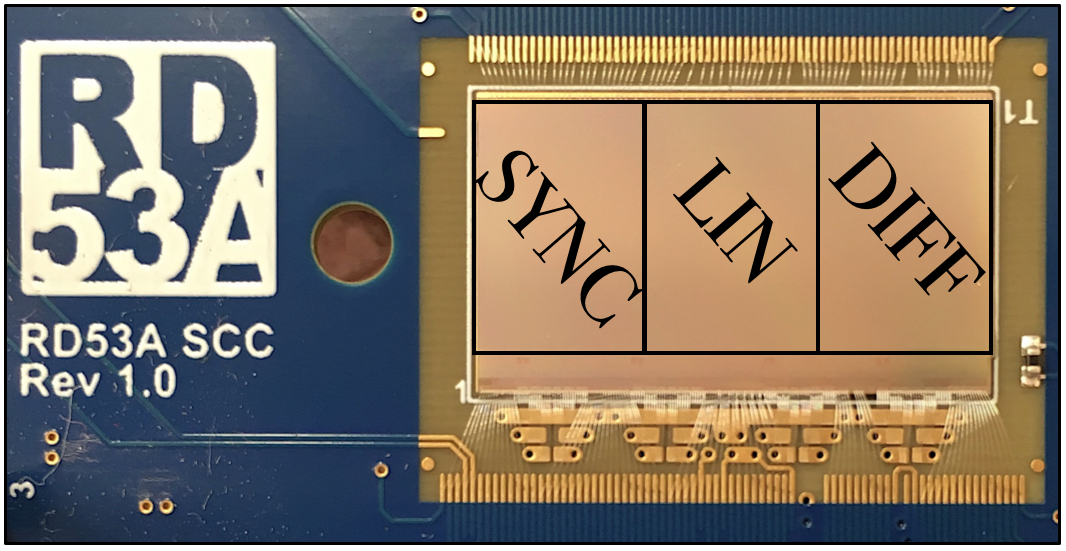}
        \caption{Photograph of the RD53A chip, wire-bonded to a test card, indicating the placement of the three \acrlong{afe}s.}
        \label{fig:rd53achip}
    \end{minipage}
\end{figure}

The RD53A \gls{afe}s are grouped by four, i.e.~$2\times2$ pixels, into analogue "islands", which are embedded in a synthesized digital “sea”, as shown in Figure~\ref{fig:ana_island}.
Three different \gls{afe} designs have been proposed within the RD53 project to explore different options for the ATLAS and \gls{cms} experiments leaving open the possibility that the experiments might make different choices.
The chip is divided horizontally into three sections, each one having one \gls{afe} design, as indicated in Figure~\ref{fig:rd53achip}. The \gls{sync} \gls{afe} is implemented between columns \num{0} and \num{127}, the \gls{lin} \gls{afe} between columns \num{128} and \num{263}, and the \gls{diff} \gls{afe} between columns \num{264} and \num{399}. It was not possible to have an equal area for all three designs because the 400-pixels wide matrix is built of \num{8x8} pixel cores~\citep{rd53a_manual}.
The three \glspl{afe} share the digital logic and the chip periphery in the RD53A chip~\citep{rd53a_manual}.
All three \gls{afe}s are based on a \gls{csa} with a feedback loop ensuring the return to baseline of the \gls{pa} output after each hit. The gain of the \gls{pa} can be chosen globally thanks to different feedback capacitors (C\textsubscript{F}) present in each \gls{afe}. The specific features of each \gls{afe} are discussed in the following paragraphs. 

\begin{figure}[p]
    \centering
    \includegraphics[width=\textwidth]{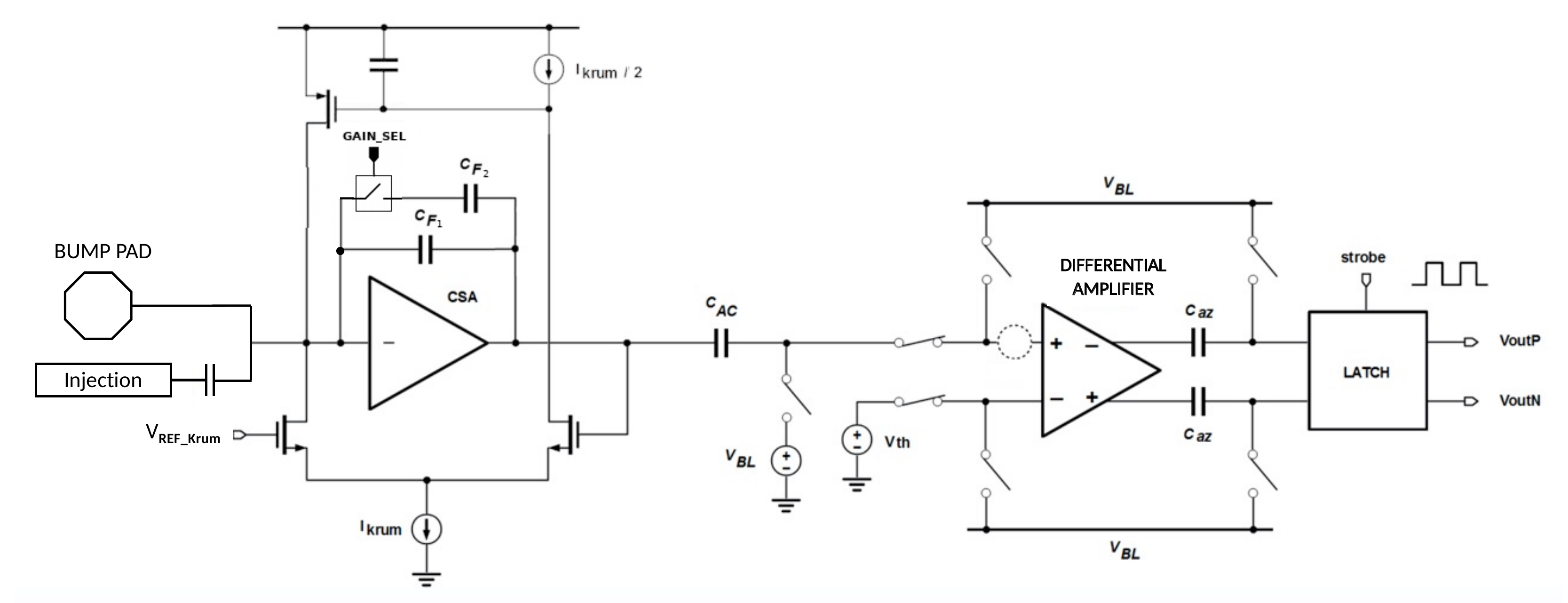}
    \caption{Schematic of the Synchronous front-end implemented in the RD53A chip~\citep{rd53a_manual}.}
    \label{fig:sync}
    \vspace{0.5cm} 
    \includegraphics[width=1.1\textwidth]{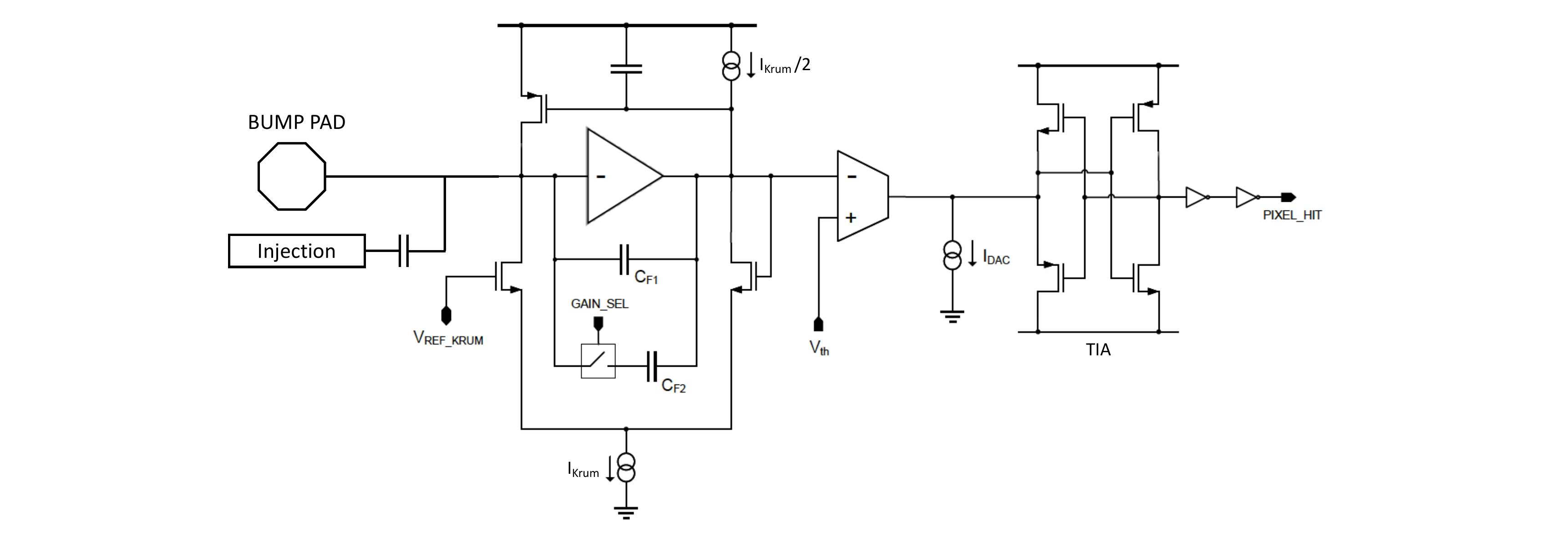}
    \caption{Schematic of the Linear front-end implemented in the RD53A chip~\citep{rd53a_manual}.}
    \label{fig:lin}
    \vspace{0.5cm} 
    \includegraphics[width=1.1\textwidth]{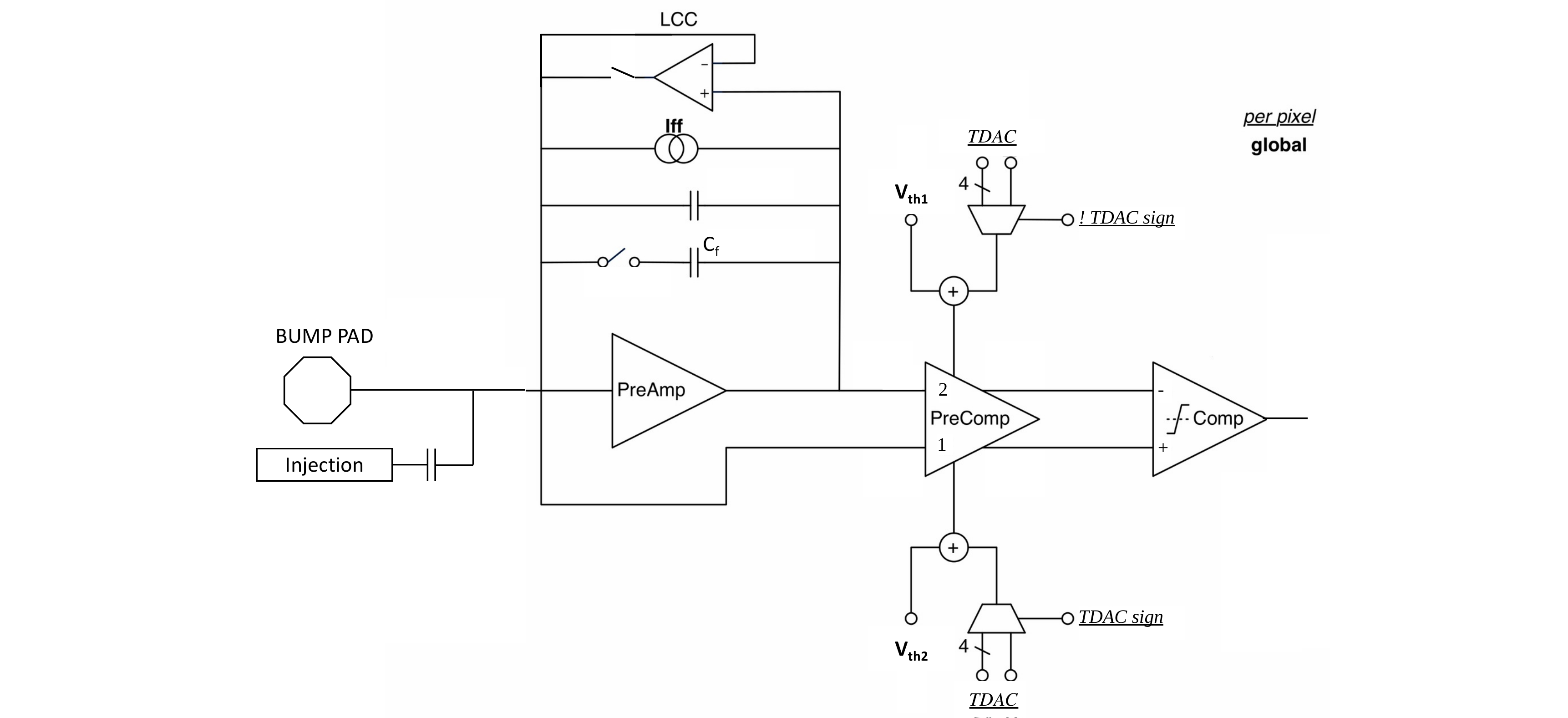}
    \caption{Schematic of the Differential front-end implemented in the RD53A chip~\citep{rd53a_manual}.}
    \label{fig:diff}
\end{figure}

\paragraph{Synchronous front-end.}
The schematic of the Synchronous front-end is shown in Figure~\ref{fig:sync}. It features a single-stage \gls{csa} with a Krummenacher feedback (I\textsubscript{Krum}, V\textsubscript{\texttt{REF\_Krum}})~\citep{krum}, which ensures both the sensor leakage current compensation and the constant current discharge of the feedback capacitor. The Krummenacher current (I\textsubscript{Krum}) drives the speed of the \gls{pa} output return to baseline.
The \gls{pa} is AC-coupled (C\textsubscript{AC}) to a synchronous discriminator composed of a \acrlong{da}, providing a further small gain, and a positive feedback latch, which performs the signal comparison with a threshold (V\textsubscript{th}) and generates the discriminator output. The latter can also be switched to a local oscillator with a selectable frequency higher than the standard \gls{lhc} clock, in order to perform a fast \gls{tot} counting.
The distinctive feature of this \gls{afe} is a so-called "auto-zero" functionality. In traditional designs, the transistor mismatch causing pixel-to-pixel variations of the threshold is compensated with a trimming \gls{dac}. In the \gls{sync} \gls{afe} instead, internal capacitors (C\textsubscript{az}) are used to compensate voltage offsets automatically. A periodic acquisition of a baseline (V\textsubscript{BL}) is required, which can be done during \gls{lhc} abort gaps~\citep{rd53a_manual, sync}. 

\paragraph{Linear front-end.}
The Linear front-end 
implements a linear pulse amplification in front of the discriminator. 
The schematic of this \gls{afe} is shown in Figure~\ref{fig:lin}. As for the \gls{sync} \gls{afe}, the \gls{pa} of the \gls{lin} \gls{afe} is based on a \gls{csa} featuring a Krummenacher feedback (I\textsubscript{Krum}, V\textsubscript{\texttt{REF\_Krum}}). The signal from the \gls{csa} is fed to a low power threshold discriminator based on current comparison, which compares the signal with the threshold (V\textsubscript{th}). It is composed of a transconductance stage followed by a \gls{tia} providing a low impedance path for fast switching. A 4-bit binary weighted trimming \gls{dac} with adjustable range (I\textsubscript{DAC}) allows for a reduction in the threshold dispersion across the pixel matrix~\citep{rd53a_manual, lin}.

\paragraph{Differential front-end.}
The \gls{pa} of the Differential front-end, shown in Figure~\ref{fig:diff}, has a continuous reset (I\textsubscript{ff}), unlike the other two designs, which use the Krummenacher feedback with constant current reset. This continuous feedback is able to prevent the input from saturation for a leakage current of up to \SI{2}{\nano\ampere}~\cite{rd53b_manual}. For higher currents, a dedicated \gls{lcc} circuit can be enabled. The \gls{lcc} is disconnected from the input when disabled, which improves the \gls{afe} stability and noise performance. The DC-coupled precomparator provides additional gain in front of the comparator and acts as a differential threshold circuit, i.e.~the global threshold is adjustable through two distributed threshold voltages (V\textsubscript{th1} and V\textsubscript{th2}) instead of one. The precomparator stage is followed by a classic time-continuous  comparator. The threshold is trimmed in each pixel using a local 5-bit trimming \gls{dac} (TDAC)~\citep{rd53a_manual}.


{The basic functionalities of the RD53A chip and each of the three \gls{afe}s were previously verified and reported~\citep{timon_hiroshima, luigi_vertex, aleksandra_vci, ennio_twepp, luigi_nima}. The objective of this work was to evaluate the three \gls{afe} designs against the \gls{cms} requirements, in terms of spurious hit rate, dead time, and radiation tolerance and to compare their performance.
A dedicated evaluation program was established and the most relevant detector performance parameters were studied. The key measurements that enabled \gls{cms} to identify the most suitable option for integration into the \gls{cms} pixel detector are presented in this paper. All presented test results were obtained with the BDAQ53 test system~\citep{bdaq53}, using the calibration injection circuit, with RD53A chips bump-bonded to sensors with rectangular pixels, i.e.~\SI[product-units = power]{100x25}{\micro\meter}, if not otherwise stated, and operated at cold temperature (T $\approx$ \SI{-10}{\celsius}), which is the lowest temperature that could be achieved with the cooling systems available for the lab setups.}
\section{CMS requirements for the analogue front-end}
\label{sec:requirements}

\begin{figure}[!b]
    \centering
    \includegraphics[width=0.49\textwidth]{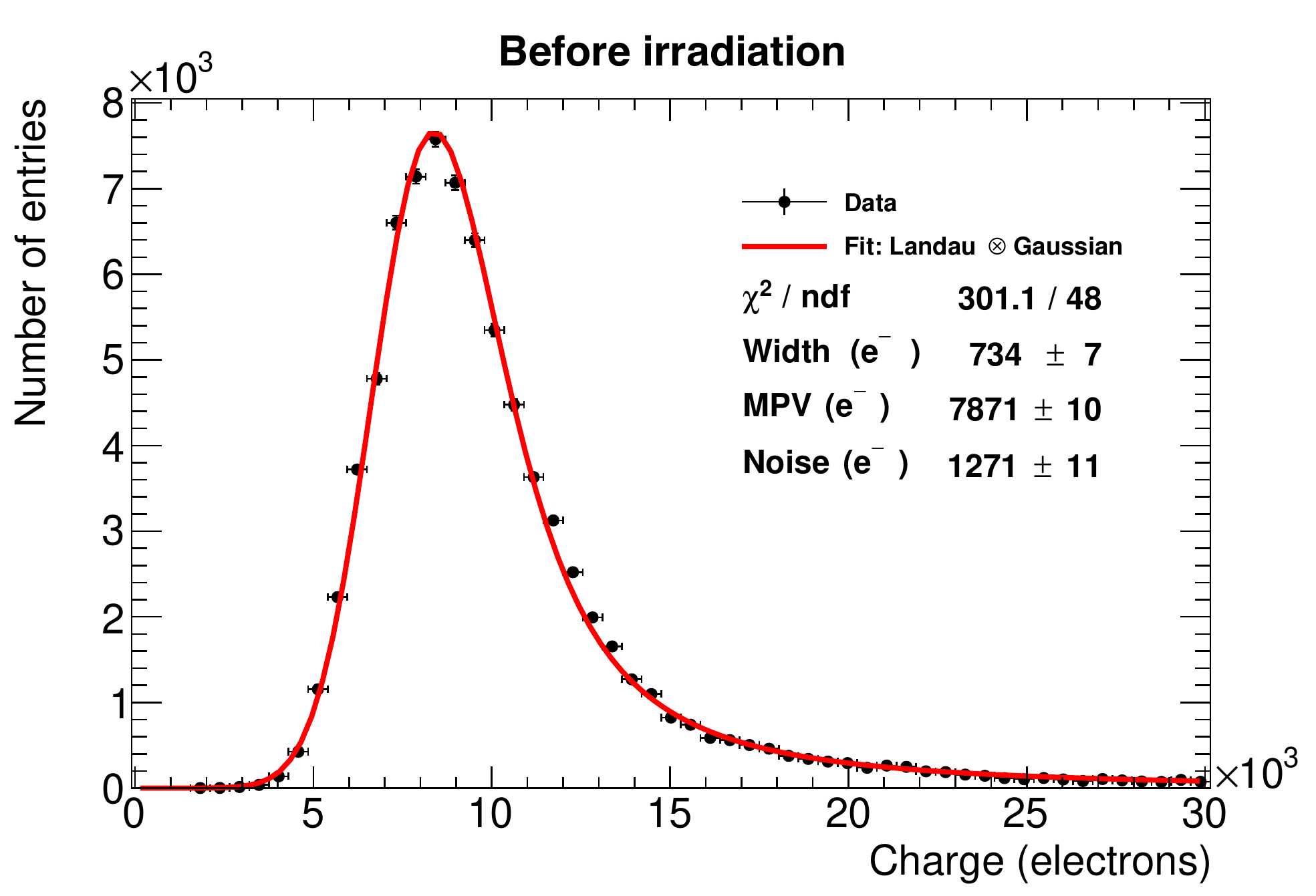}
    \includegraphics[width=0.49\textwidth]{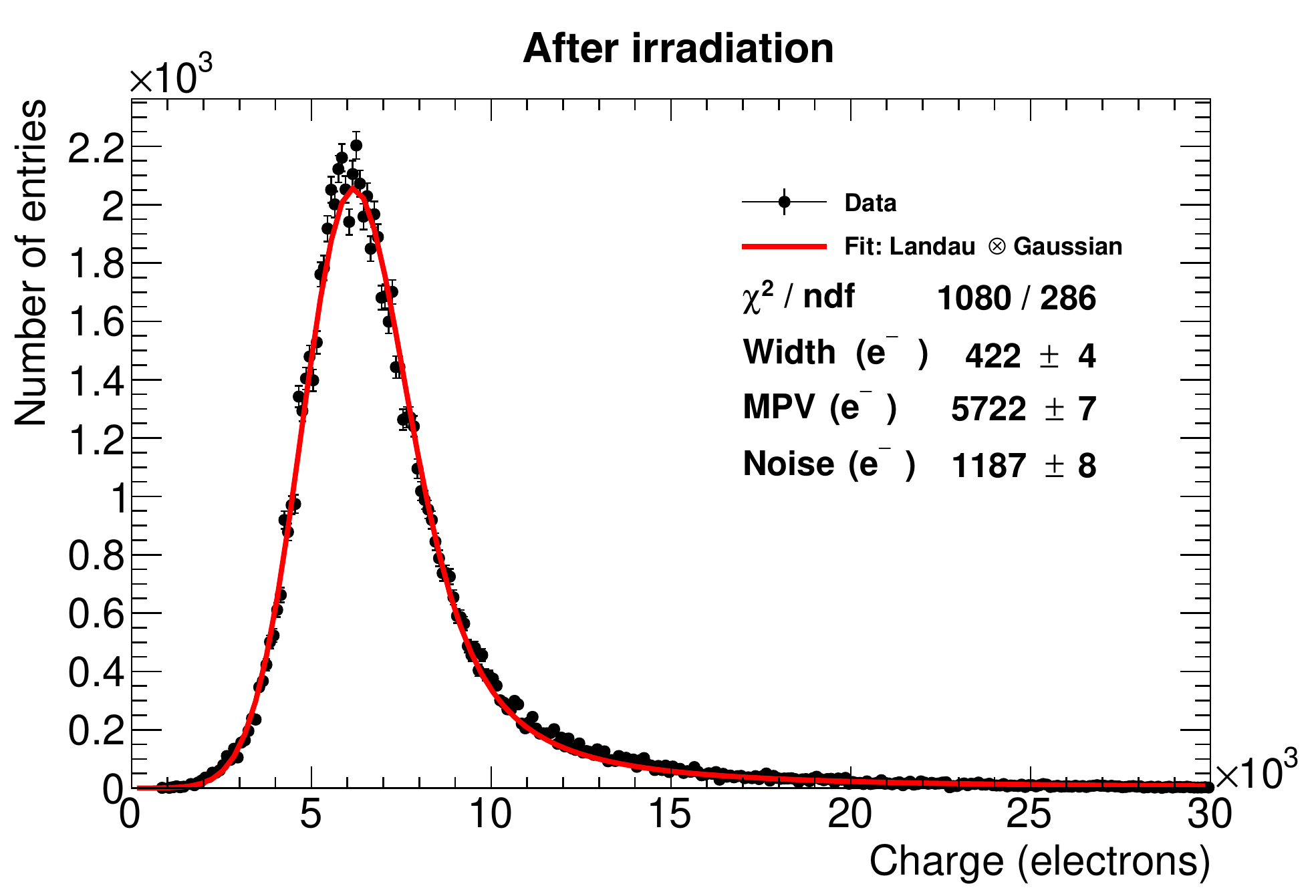}
    \caption{Test beam measurement of the collected charge before (left) and after (right) irradiation to \SI{1.2e15}{n_{eq}\per\centi\meter\squared}, using single pixel clusters, in a \SI{130}{\micro\meter} thick pixel sensor with \SI[product-units = power]{100x150}{\micro\meter} pixels. The red line represents a fit to a Landau distribution convoluted with a Gaussian~\cite{p2_tdr}.}
    \label{fig:mip}
\end{figure}

The first step towards the choice of the \gls{afe} for the \gls{cms} final chip was the establishment of the evaluation criteria. The most relevant detector parameters were used to derive the following \gls{cms} requirements:

\paragraph{Optimal threshold.}
The new \gls{cms} readout chip will feature \SI[product-units = power]{50x50}{\micro\meter} pixels, while the pixel size is \SI[product-units = power]{100x150}{\micro\meter} in the present \gls{cms} pixel detector~\citep{p1_tdr}. The readout chip can be bump-bonded either to sensors with square pixels of the same size or to rectangular pixels of \SI[product-units = power]{100x25}{\micro\meter}, thanks to electrode routing in the sensor~\citep{p2_tdr}. Silicon sensors with a thickness of \SI{150}{\micro\meter} will be used. This is about half the thickness of the current \SI{285}{\micro\meter} thick sensors~\citep{p1_tdr}. The main advantage of thin sensors is better radiation tolerance, but the collected signal charge is smaller. The charge distribution obtained with \SI{120}{\giga\electronvolt} protons from a test beam collected in a \SI{130}{\micro\meter} thick sensor with \SI[product-units = power]{100x150}{\micro\meter} pixels is shown in Figure~\ref{fig:mip}. The \gls{mpv} is about \num{7900}~e${^{-}}$ before irradiation. While this number is about \SI{10}{\percent} lower than the expectation, it is well compatible with it within the measurement uncertainties (e.g. due to the charge calibration). The \gls{mpv} decreases by about \num{2000}~e${^{-}}$ after irradiation to \SI{1.2e15}{n_{eq}\per\centi\meter\squared}~\citep{p2_tdr}. Based on the expected signal, a detection threshold of \num{1000}~e${^{-}}$ is required by \gls{cms} for the innermost layer of the \gls{it} to ensure sufficient detection efficiency, especially with irradiated sensors. A threshold of \num{1200}~e${^{-}}$ is sufficient for the outer layers of the detector, where the fluence is lower. 

The number of pixels hit increases with the incidence angle of the particle, giving clusters with large hit multiplicity in particular in the high-$\eta$ part of the barrel. In this specific part of the detector, the charge collection path in a pixel is similar to the pixel dimension in the $z$ direction, hence $\gtrsim$~\SI{50}{\micro\meter} for square pixels and $\gtrsim$~\SI{100}{\micro\meter} for rectangular pixels, to be compared with a charge collection path of \SI{150}{\micro\meter} at normal incidence. For this reason, in the high-$\eta$ region of the barrel square pixels are disfavoured, being more prone to remain below threshold, notably after irradiation.

\paragraph{Radiation tolerance.}
The \gls{it} is the \gls{cms} subdetector closest to the \gls{lhc} interaction point and therefore it is exposed to the highest radiation levels. Two scenarios are envisaged for the HL-LHC: in the "nominal" scenario, the accelerator would deliver a maximum of \num{140} proton-proton (pp) collisions per bunch crossing, to reach a total integrated luminosity of \SI{3000}{\per\femto\barn} by the end of the physics program. In the "ultimate" scenario, the number of pp collisions per bunch crossing would be pushed up to 200, reaching an integrated luminosity of \SI{4000}{\per\femto\barn}.
A fluence reaching \SI{2.6e16}{n_{eq}\per\centi\meter\squared} and a \gls{tid} up to \SI{1.4}{\giga\rad} are expected in the innermost layer in the nominal scenario, while the figures would scale up to \SI{3.4e16}{n_{eq}\per\centi\meter\squared} and \SI{1.9}{\giga\rad} in the ultimate scenario.
The RD53A chip was designed to withstand a \gls{tid} of at least \SI{500}{\mega\rad} and an average leakage current up to \SI{10}{\nano\ampere\per pixel}~\citep{rd53a_specs}. However, with this specification the radiation levels expected in \gls{cms}, reaching \SI{1.9}{\giga\rad} in the ultimate luminosity scenario, would imply a replacement of the innermost layer of the \gls{it} barrel after every two years of operation. The \gls{cms} Collaboration aims for a single replacement of the innermost layer during the ten-year lifetime of the detector, hence a higher radiation tolerance is necessary.

\begin{figure}[b]
    \centering
    \includegraphics[width=0.6\textwidth]{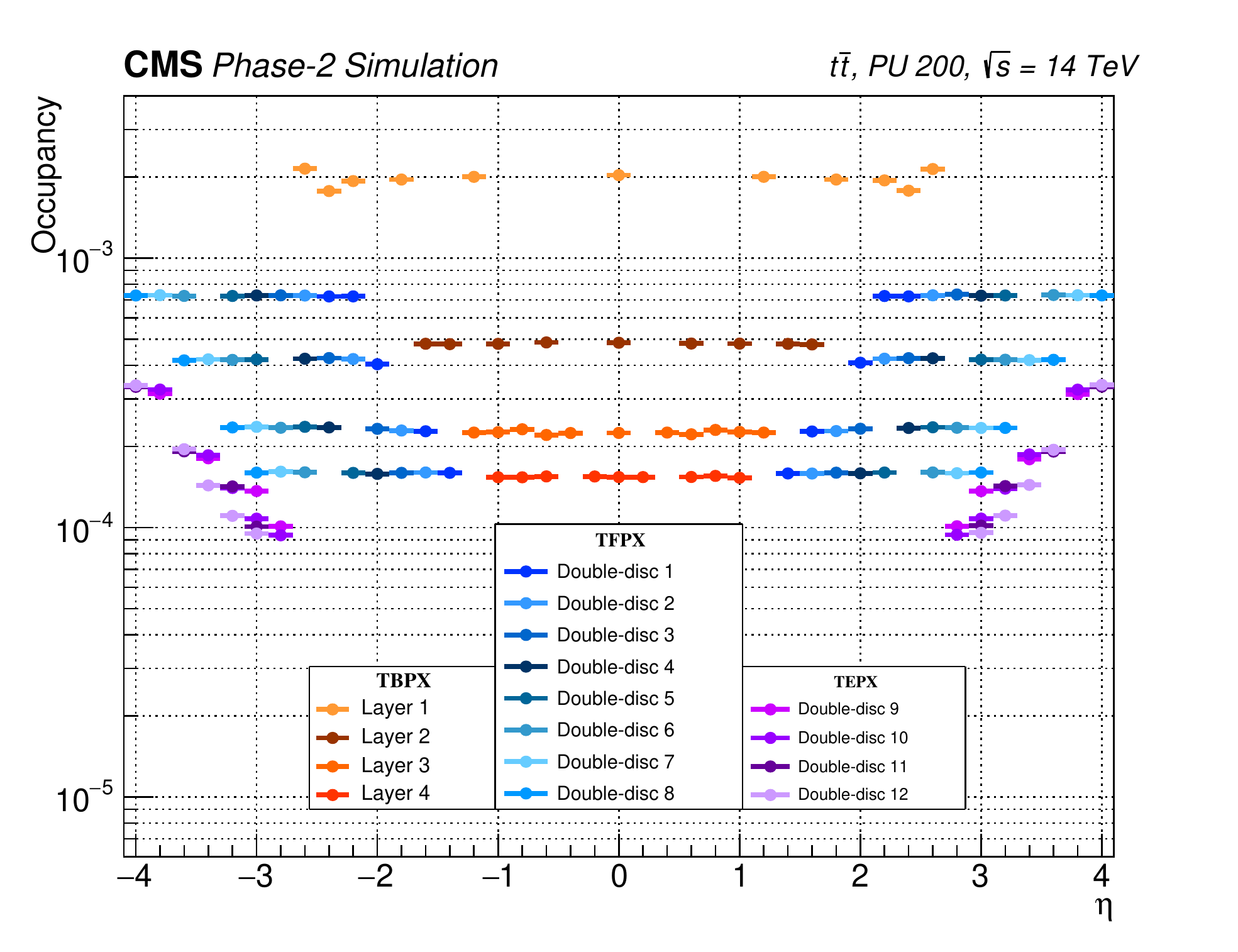}
    \caption{Simulation of the hit occupancy as a function of pseudorapidity for all layers and double-discs of the \gls{it} for simulated top quark pair production events with a pileup of 200 events~\citep{p2_tdr}.}
    \label{fig:occ}
\end{figure}

\paragraph{Noise occupancy.}
For a stable operation at low threshold it is important to minimize the front-end noise to have an acceptable fraction of spurious hits in the data. Single pixels that are too noisy can be disabled, to keep the overall noise occupancy low, but their fraction must be low in order not to significantly affect the detector efficiency.
Based on the occupancy simulation for different parts of the detector, shown in Figure~\ref{fig:occ}, the average noise occupancy of the new front-end is required to be below \num{e-6}, i.e.~two orders of magnitude below the lowest expected occupancy.

\paragraph{Dead time.} \gls{cms} requires a maximum dead time of \SI{1}{\percent} in the innermost layer of the \gls{it} barrel to ensure high detection efficiency even at the highest expected hit rate. This requirement translates to a maximum efficiency loss of \SI{1}{\percent} at maximum hit rate caused by the total dead time (digital + analogue). The dead time in the RD53A chip has a minor contribution from the digital buffering and a major contribution from the \gls{csa} of the \gls{afe}. While the digital contribution is due to the limited hit buffer size and cannot be reduced with the chip settings, the \gls{afe} dead time depends on the \gls{tot} response calibration. The \gls{tot} response to a given input charge can be set in the chip to a certain number of TOT\textsubscript{40} units (one TOT\textsubscript{40} unit corresponds to one \SI{40}{\mega\hertz} clock cycle, i.e.~to~\SI{25}{\nano\second}). The charge resolution is obtained by dividing the input charge by the corresponding number of clock cycles and can therefore be expressed in e${^{-}}$/TOT\textsubscript{40} units.

\begin{figure}[!b]
    \centering
    \includegraphics[width=0.45\textwidth]{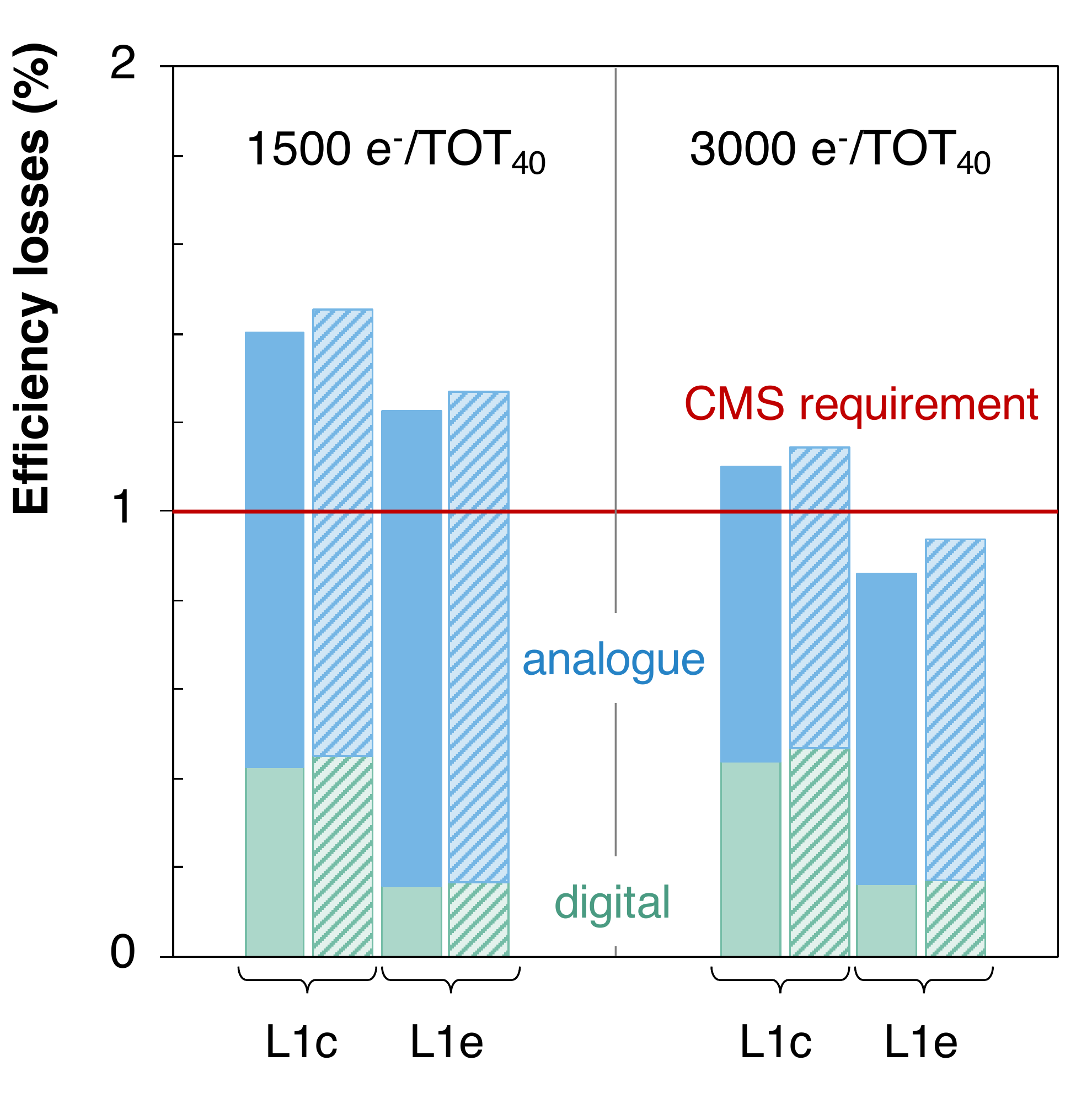}
    \caption{Hit efficiency losses due to the digital buffering (green) and analogue dead time (blue) simulated in the 200 pileup scenario for two charge resolutions: \num{1500}~e${^{-}}$/TOT$_{40}$ and \num{3000}~e${^{-}}$/TOT$_{40}$. The simulation was done for the centre (c) and edge (e) of the innermost layer (L1) of the \gls{it} barrel and for two pixel geometries: the solid bins represent the \SI[product-units = power]{100x25}{\micro\meter} pixels and hashed bins represent the \SI[product-units = power]{50x50}{\micro\meter} pixels. The red line represents the CMS requirement.}
    \label{fig:dt-simulation}
\end{figure}

A Monte Carlo simulation of hit efficiency losses due to the digital, analogue, and total dead time is shown in Figure~\ref{fig:dt-simulation} for two charge resolutions: \num{1500}~e${^{-}}$/TOT$_{40}$ and \num{3000}~e${^{-}}$/TOT$_{40}$. The simulation was performed for two pixel module positions in the innermost layer of the \gls{it} barrel: the centre ($z=0$), denoted L1c, and the edge, denoted L1e.
For each position, both pixel geometries were simulated. The rectangular pixels are represented with solid bins and the square pixels with hashed bins. The square pixels have a slightly higher inefficiency. 
As expected, the \gls{tot} charge resolution has no influence on the digital dead time, and the hit losses caused by the \gls{afe} are smaller with the coarser charge resolution of \num{3000}~e${^{-}}$/TOT$_{40}$. 
The efficiency losses are higher in the centre making the dead time requirement difficult to meet. With the charge resolution of \num{1500}~e${^{-}}$/TOT$_{40}$ the requirement is not satisfied in any of the two module positions, while with \num{3000}~e${^{-}}$/TOT$_{40}$ the requirement is satisfied on average. The hit efficiency losses are slightly above the requirement in the centre and slightly below at the edge. The charge resolution of \num{3000}~e${^{-}}$/TOT$_{40}$ was therefore taken as the \gls{tot} calibration requirement for the \gls{afe} evaluation.

\begin{figure}[htb]
    \centering
    \begin{subfigure}{0.49\textwidth}
        \includegraphics[width=\textwidth]{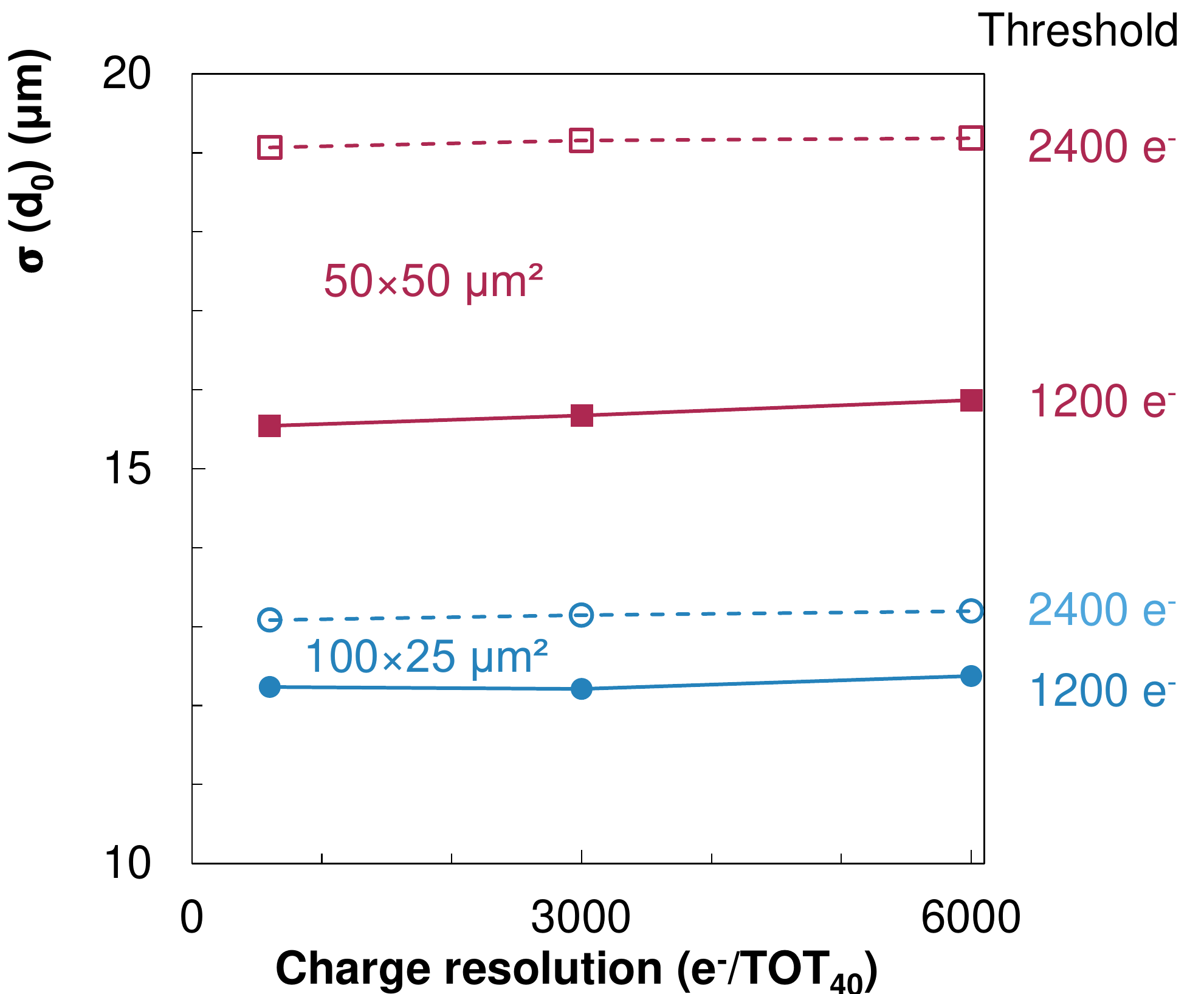}
    \end{subfigure}
    \hfill
    \begin{subfigure}{0.49\textwidth}
        \includegraphics[width=\textwidth]{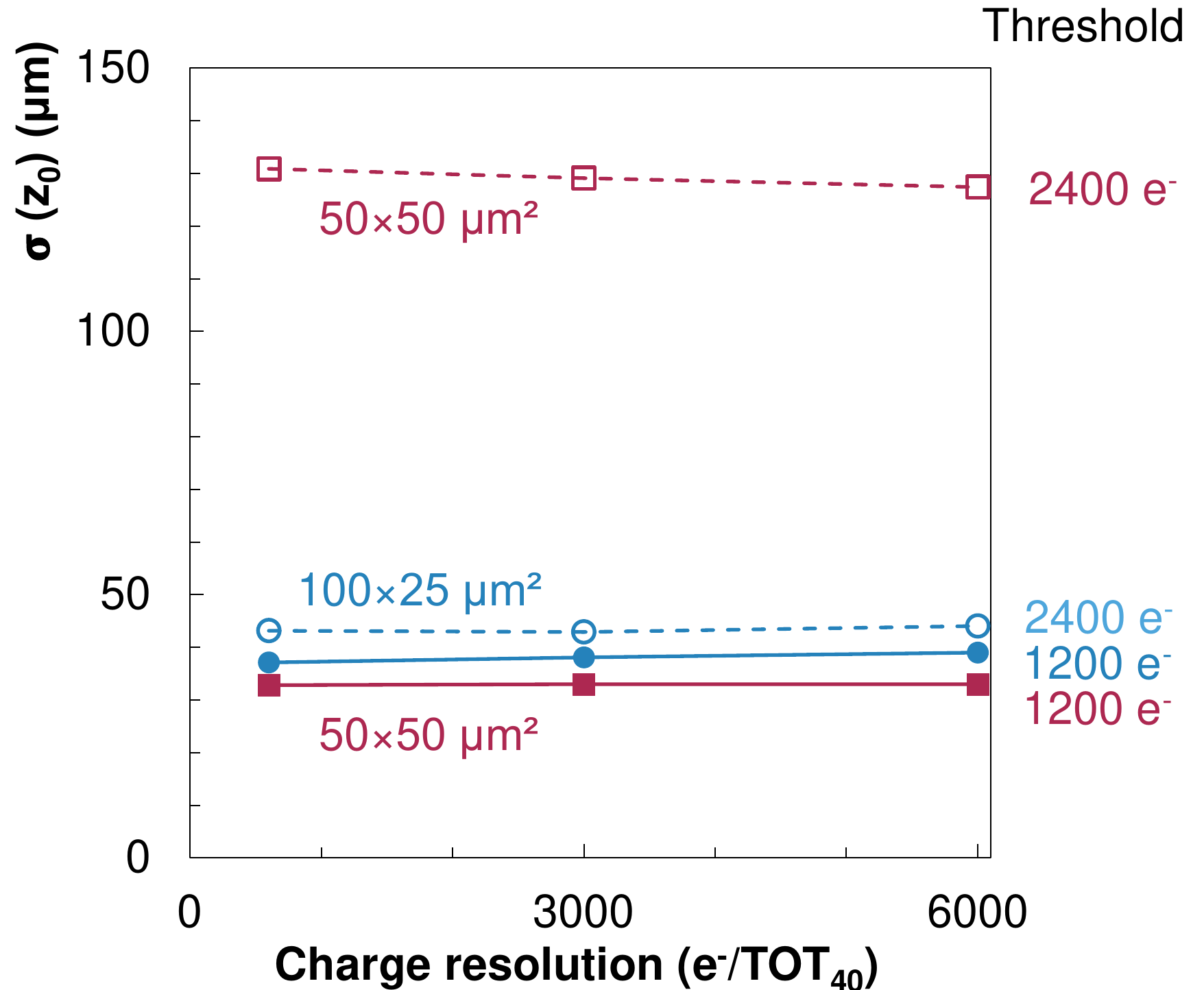}
    \end{subfigure}  
    \caption{Influence of the charge resolution on the transverse (left) and longitudinal (right) impact parameter resolution obtained from simulation. The \SI[product-units = power]{100x25}{\micro\meter} pixels are represented in blue and the \SI[product-units = power]{50x50}{\micro\meter} pixels are represented in red. The full markers and solid lines indicate the threshold of \SI{1200}{e^-} and the open markers and dashed lines indicate the threshold of \SI{2400}{e^-}.}
    \label{fig:dt-tot-resolution}
\end{figure}

The impact of charge resolution on tracking performance was also evaluated.
Simulation of the tracking performance for the reconstruction of single muons with a transverse momentum of \SI{10}{\giga\electronvolt} was performed with planar \SI{150}{\micro\meter}-thick sensors, with both sensor pixel geometries and two different thresholds: \num{1200}~e${^{-}}$ and \num{2400}~e${^{-}}$.
The resolution on the transverse ($d$\textsubscript{0}) and longitudinal ($z$\textsubscript{0}) impact parameters, denoted $\sigma(d$\textsubscript{0}) and $\sigma(z$\textsubscript{0}) respectively, integrated over the full $\eta$ range are shown in Figure~\ref{fig:dt-tot-resolution} for three charge resolutions: \num{600}~e${^{-}}$/TOT$_{40}$, \num{3000}~e${^{-}}$/TOT$_{40}$ and \num{6000}~e${^{-}}$/TOT$_{40}$.
The impact parameter resolution deteriorates for a higher threshold and appears to be insensitive to the charge resolution. Since a higher charge resolution does not affect the tracking performance, a charge resolution of \num{3000}~e${^{-}}$/TOT$_{40}$ was taken as the baseline calibration for the inner regions of the Inner Tracker.

\section{Equalization of threshold dispersion}
\label{sec:rad}

\begin{figure}[ht]
    \centering
    \begin{subfigure}{0.37\textwidth}
        \centering
        \captionsetup{justification=centering}
        \includegraphics[width=\textwidth]{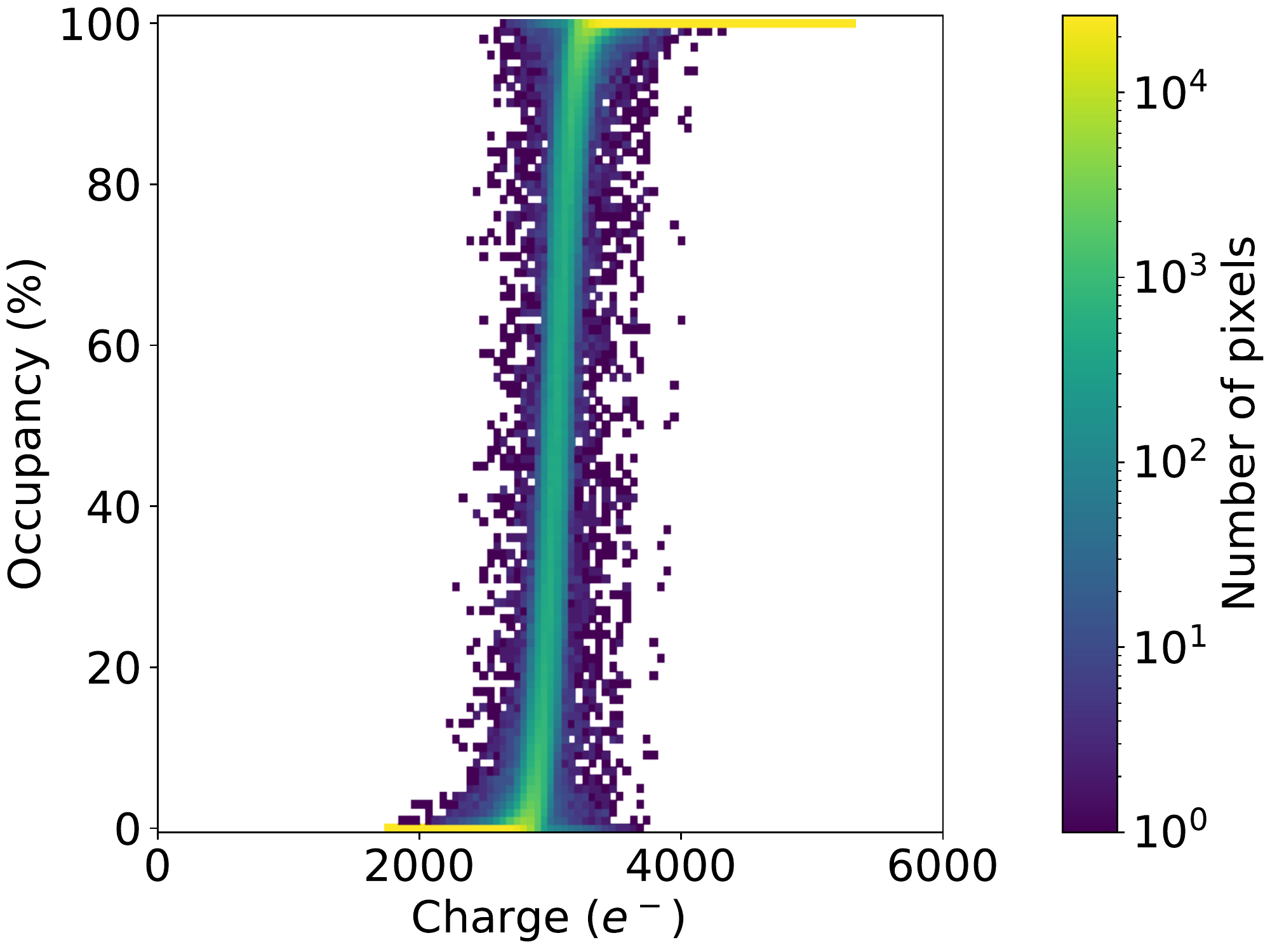}
        \caption{S-curves.\newline}
        \label{fig:scurves}
    \end{subfigure}
    \hfill
    \begin{subfigure}{0.29\textwidth}
        \centering
        \captionsetup{justification=centering}
        \includegraphics[width=\textwidth]{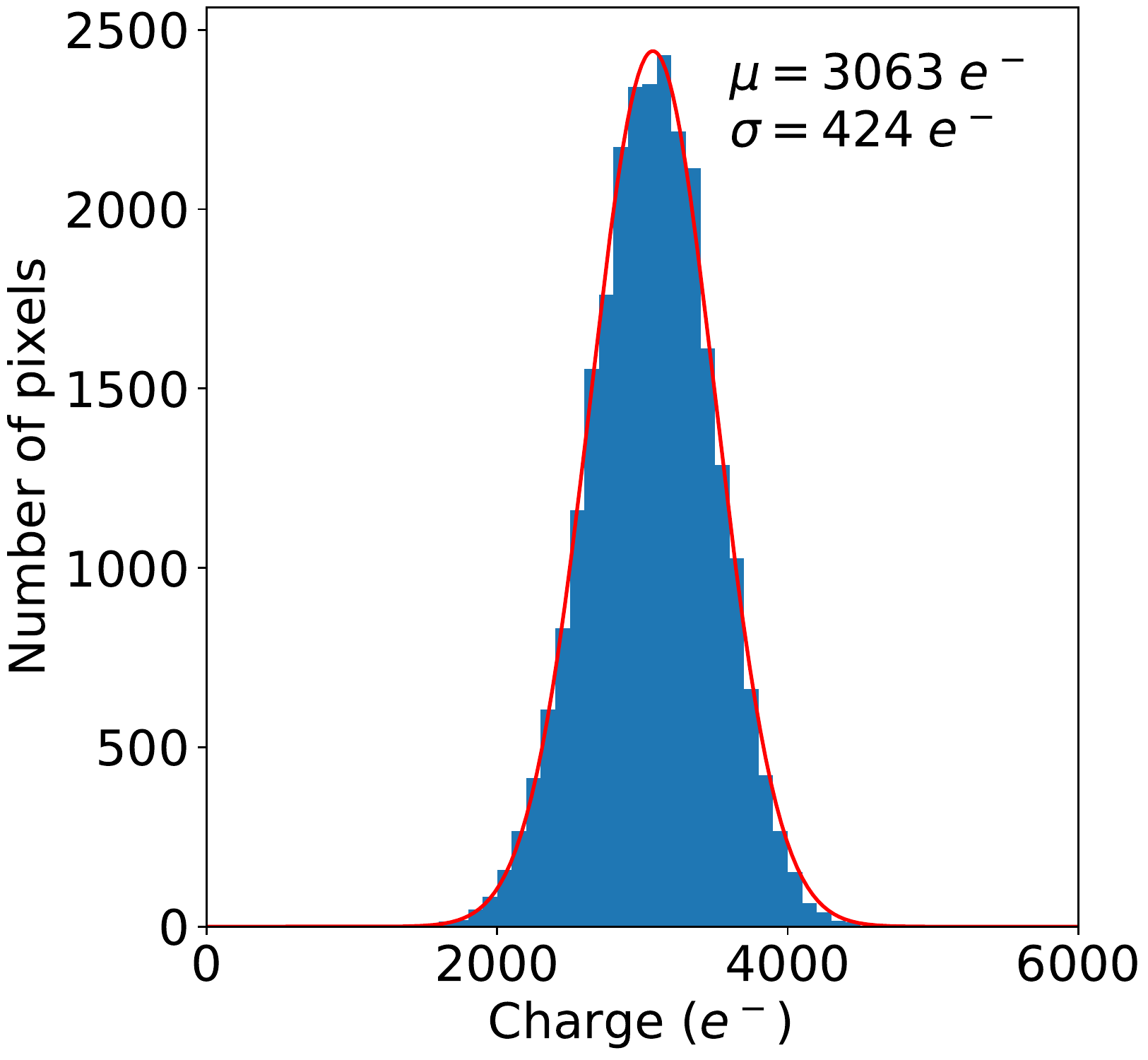}
        \caption{Untuned threshold distribution.}
        \label{fig:uthrdist}
    \end{subfigure}
    \hfill
    \begin{subfigure}{0.29\textwidth}
        \centering
        \captionsetup{justification=centering}
        \includegraphics[width=\textwidth]{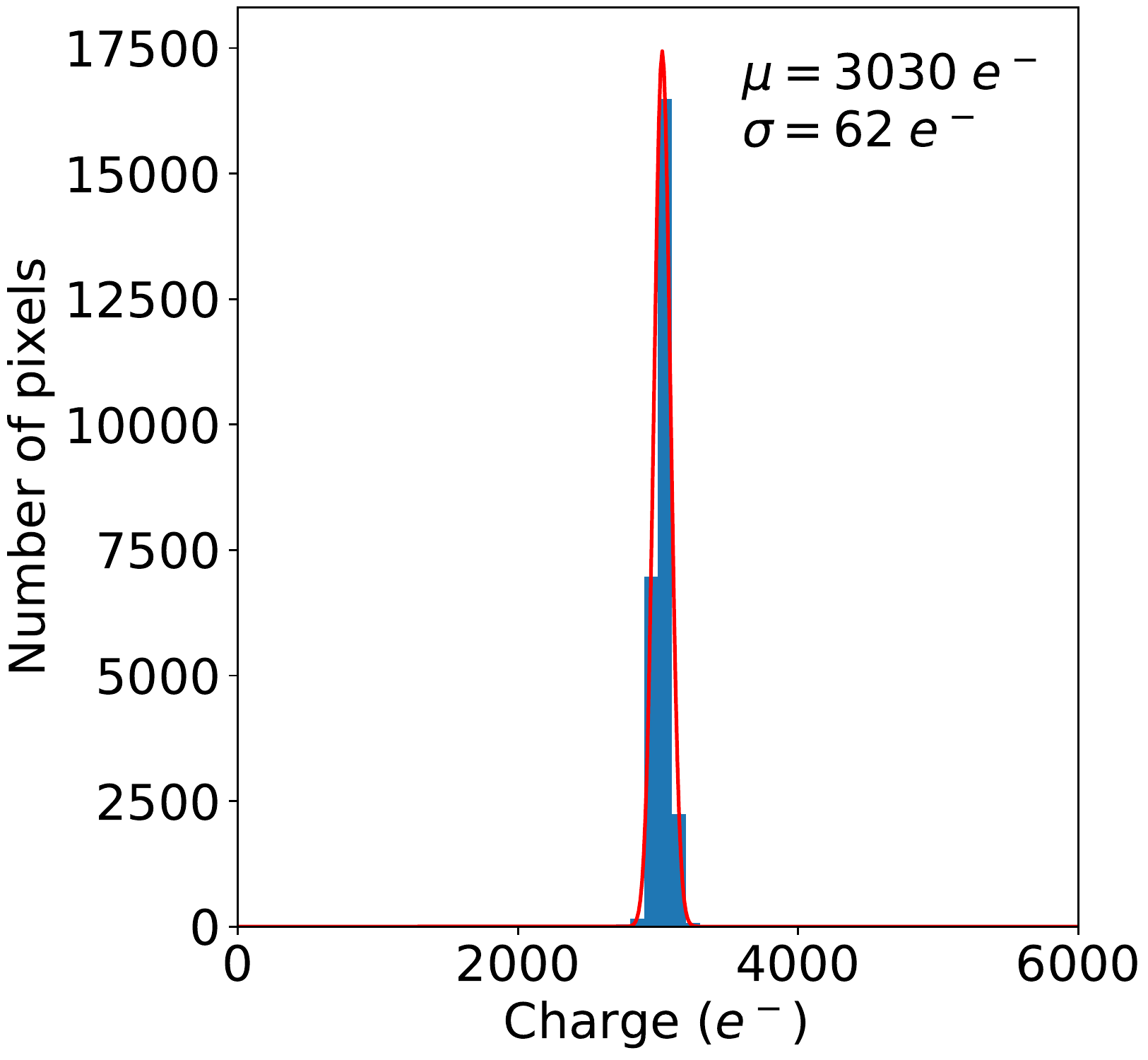}
        \caption{Tuned threshold distribution.}
        \label{fig:tthrdist}
    \end{subfigure}
    \label{fig:thrdist}
    \caption{S-curves (a) and threshold distribution before (b) and after (c) tuning obtained with all pixels of the \gls{lin} \gls{afe} of one RD53A chip. The red lines in the threshold distributions represent fits to Gaussian distributions and the mean ($\mu$) and width ($\sigma$) of the fit functions are given.}
\end{figure}

The calibration injection circuit is used to inject a range of charges to measure the threshold of each pixel and the threshold dispersion across the matrix. The occupancy versus charge of a pixel is a sigmoid from \num{0} to \SI{100}{\percent} occupancy, commonly called an S-curve. An example of an S-curve plot including more than \num{26000} pixels is shown in Figure~\ref{fig:scurves}. 
The calibration charge at which \SI{50}{\percent} occupancy is reached is taken as a measurement of the charge equivalent of the threshold of each pixel. 
The mean value of the pixel threshold distribution represents the global threshold and the \gls{rms} is the threshold dispersion.
Typically, pixel-to-pixel variations result in a threshold dispersion of several hundred electrons, which can be reduced to less than hundred electrons after setting optimal trim bits for each pixel with a dedicated tuning algorithm. Examples of untuned and tuned threshold distributions are shown in Figure~\ref{fig:uthrdist} and Figure~\ref{fig:tthrdist}, respectively.

\begin{figure}[t]
    \centering
    \begin{subfigure}{0.32\textwidth}
        \centering
        \includegraphics[width=\textwidth]{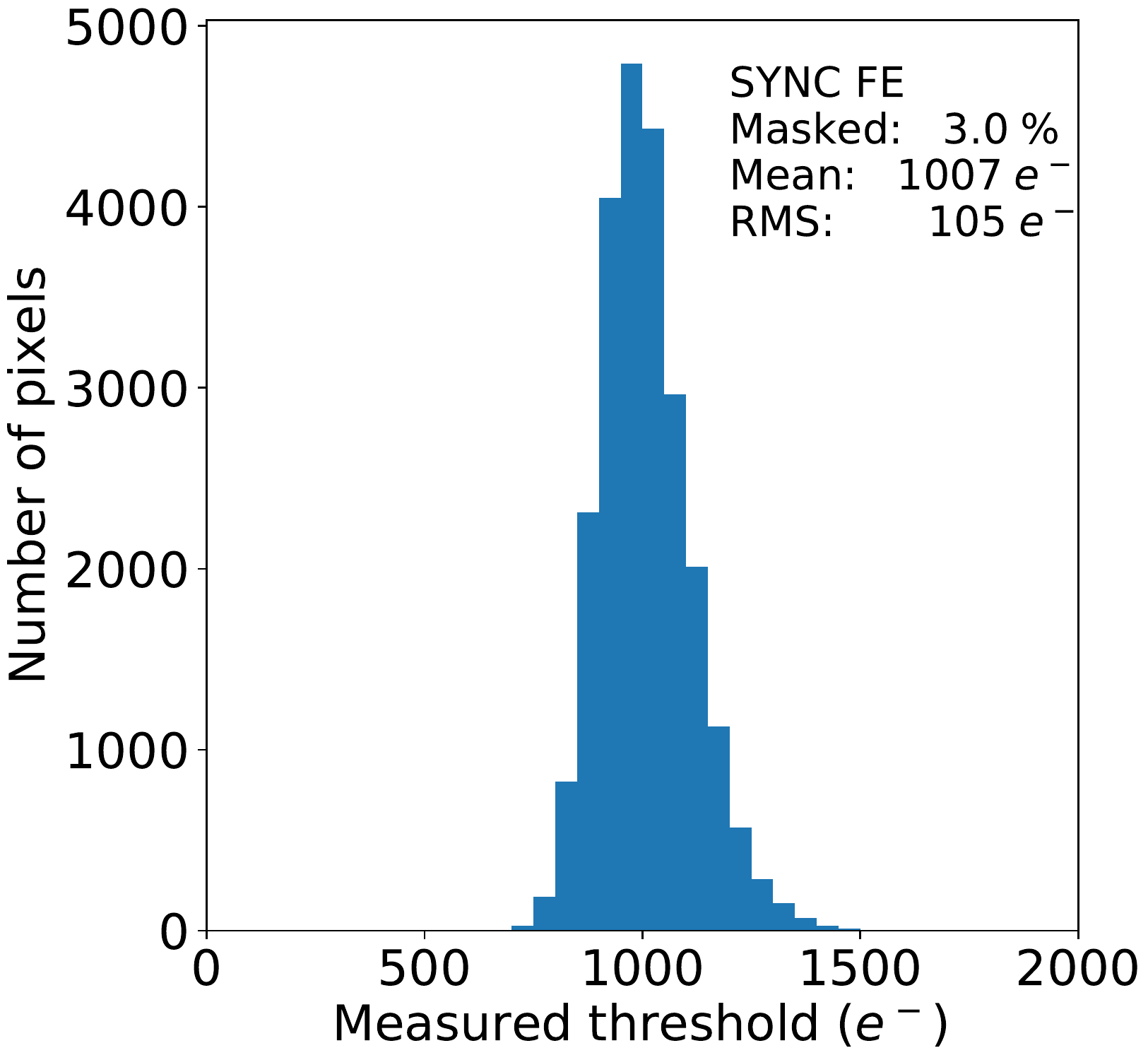}
        \caption{SYNC AFE.}
    \end{subfigure}
    \hfill
    \begin{subfigure}{0.32\textwidth}
        \centering
        \includegraphics[width=\textwidth]{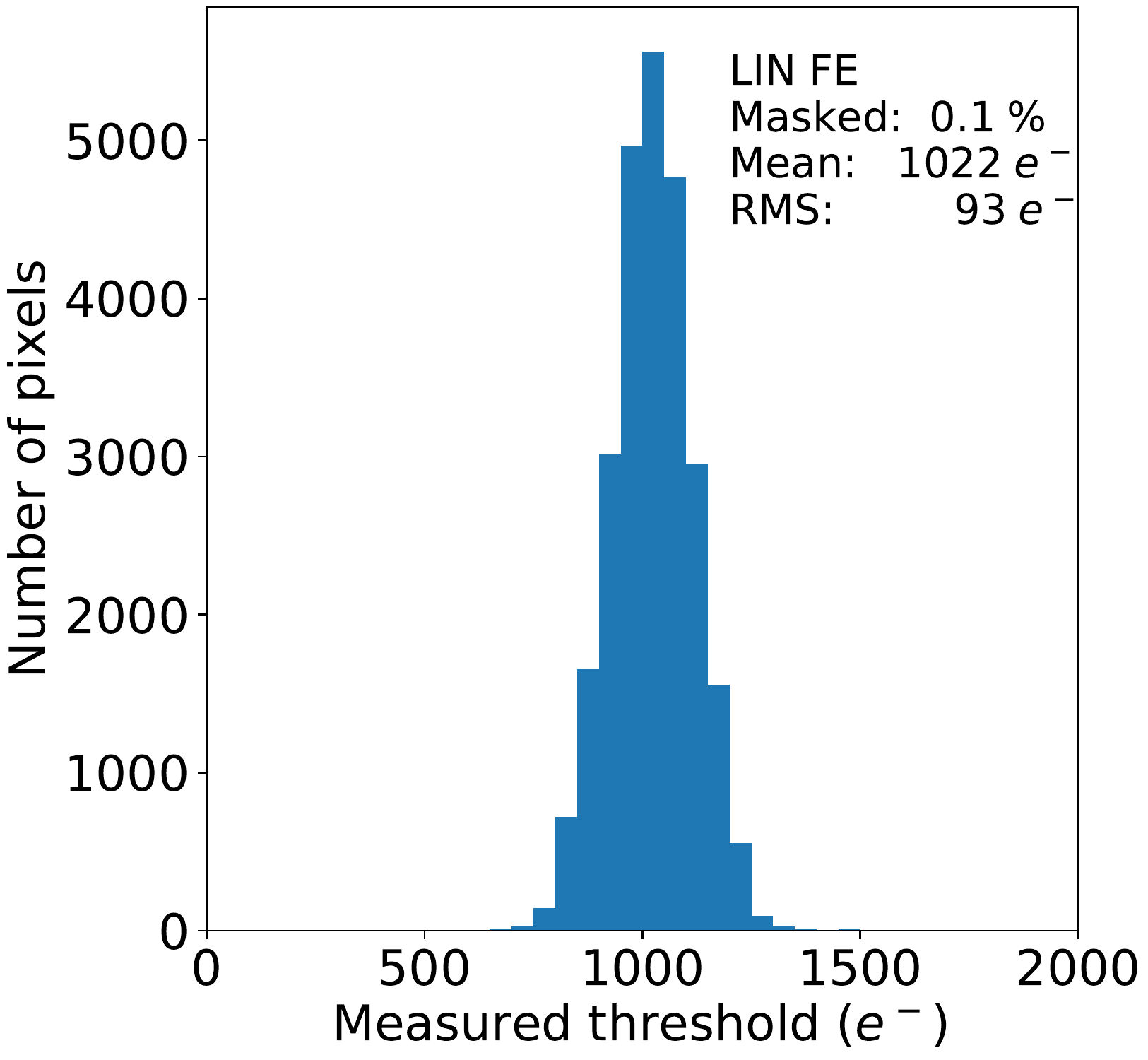}
        \caption{LIN AFE.}
    \end{subfigure}
    \hfill
    \begin{subfigure}{0.32\textwidth}
        \centering
        \includegraphics[width=\textwidth]{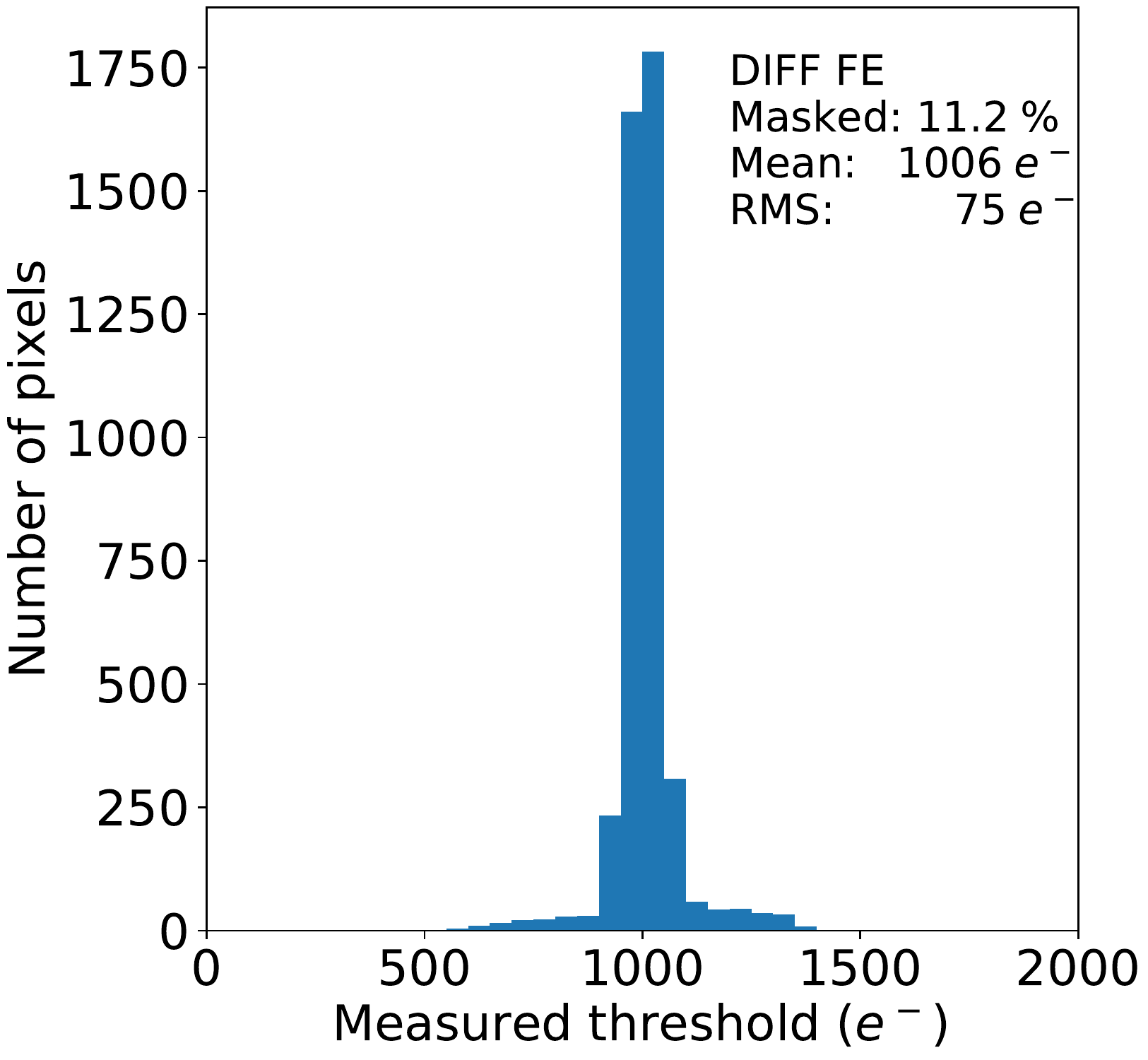}
        \caption{DIFF AFE.}
    \end{subfigure}
    \caption{Threshold distributions of the \gls{sync} \gls{afe} (a), the \gls{lin} \gls{afe} (b), and the \gls{diff} \gls{afe}~(c) obtained with an RD53A chip after irradiation. The mean and \gls{rms} were calculated using all non-masked pixels in each \gls{afe}.}
    \label{fig:irrad_3afes_thr}
\end{figure}

The threshold tuning capability of the three \gls{afe} designs was tested and proven to be functional in many samples, some of which were irradiated and re-evaluated afterwards. An assembly of an RD53A chip and a sensor with \SI[product-units = power]{50x50}{\micro\meter\squared} pixels was irradiated at Karlsruhe Institute of Technology~\citep{kit} with \SI{23}{\mega\electronvolt} protons up to a fluence of \SI{3e15}{n_{eq}\per\centi\meter\squared}, corresponding to a \gls{tid} reaching \SI{350}{\mega\rad}. The RD53A chip was not powered during irradiation. The sample was irradiated at room temperature and maintained at cold temperature after irradiation to avoid annealing. It was tested at \SI{-10}{\celsius} in a dry environment. The sensor bias voltage was adjusted to reach an average leakage current of \SI{10}{\nano\ampere}$/$pixel, which is the maximum specified value for the RD53A chip \cite{rd53a_specs}.
The pixels with a noise occupancy higher than \num{e-4} were considered noisy and masked, based on the lowest hit occupancy expected from the simulation presented in Figure~\ref{fig:occ}. The remaining pixels were tuned to a threshold of \num{1000}~e${^{-}}$. Pixels with an anomalously high threshold that could not be adjusted with the range of the trim bits were masked in the tuning procedure. The threshold distributions of the three \gls{afe}s after tuning are shown in Figure~\ref{fig:irrad_3afes_thr}.

\begin{figure}[t]
    \centering
    \includegraphics[width=0.33\textwidth]{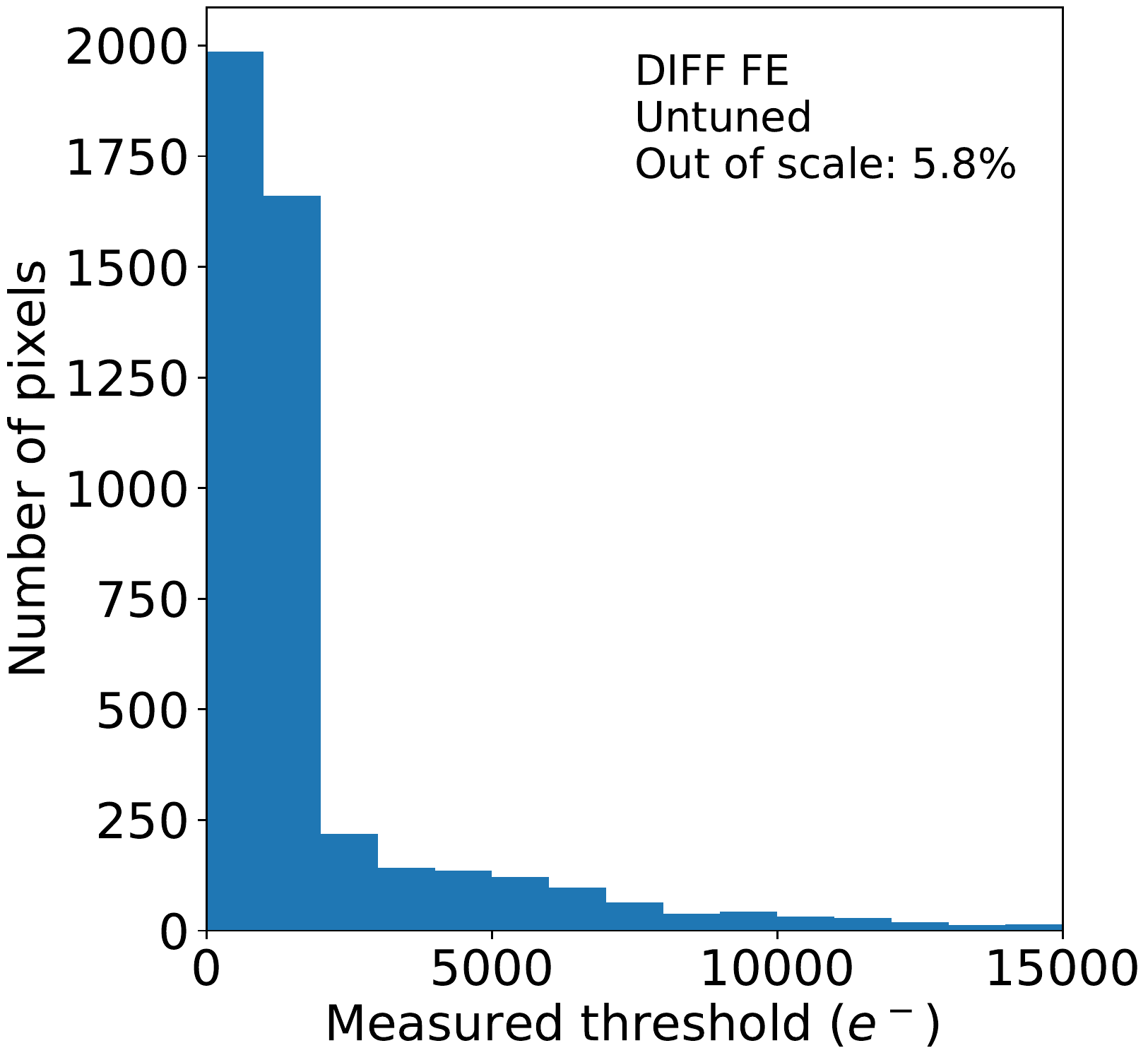}
    \caption{Untuned threshold distribution of the \gls{diff} \gls{afe} in the irradiated RD53A chip.}
    \label{fig:irrad_diff_untuned}
\end{figure}

All three \gls{afe}s were functional after irradiation and could reach the required threshold with a threshold dispersion of about \num{100}~e${^{-}}$. The threshold tuning of the \gls{lin} \gls{afe} worked well and only \SI{0.1}{\percent} of pixels were masked. The auto-zeroing in the \gls{sync} \gls{afe} worked well too, however the leakage current caused a higher noise in this front-end and \SI{3}{\percent} of the pixels were masked. The threshold distribution of the \gls{diff} \gls{afe} features a narrow core and long tails, and  \SI{11.2}{\percent} of pixels were masked.
The large fraction of masked pixels in the \gls{diff} \gls{afe} is the consequence of a long tail in the untuned threshold distribution, shown in Figure~\ref{fig:irrad_diff_untuned}. Pixels with a too high threshold cannot be tuned to the desired threshold value because the range covered by the trim bits is a global setting for the whole chip. As the threshold dispersion depends on several \gls{afe} parameters, many parameter combinations were tried to mitigate the problem, and the \SI{11.2}{\percent} of masked pixels was the best result that could be achieved with this irradiated sample.

This study triggered an investigation and the design team discovered that the combined effect of irradiation and cold temperature resulted in a PMOS threshold increase in the \gls{diff} precomparator, resulting in a small voltage margin.
Simulations showed that the voltage margin is smaller at cold temperature and decreases with irradiation, reaching a value close to zero for the \gls{diff} \gls{afe} design implemented in the RD53A chip after irradiation to \SI{200}{\mega\rad}, which explains the problematic threshold tuning observed after irradiation to \SI{350}{\mega\rad}.
A design improvement of the \gls{diff} precomparator was proposed and simulated, obtaining an extension of the expected operation range up to \SI{500}{\mega\rad}, although this radiation level is still low compared to the dose expected in the \gls{cms} detector. With such operation range, replacements of the innermost layer would be required every two years once the ultimate luminosity is reached, while \gls{cms} is aiming at a single replacement during the whole high-luminosity program. For this reason the attention turned to the other two \gls{afe}s, which seem promising candidates for a higher radiation tolerance.
    
\section{Noise evaluation}
\label{sec:noise}

\begin{figure}[t]
    \centering
    \begin{subfigure}{0.49\textwidth}
        \centering
        \includegraphics[width=\textwidth]{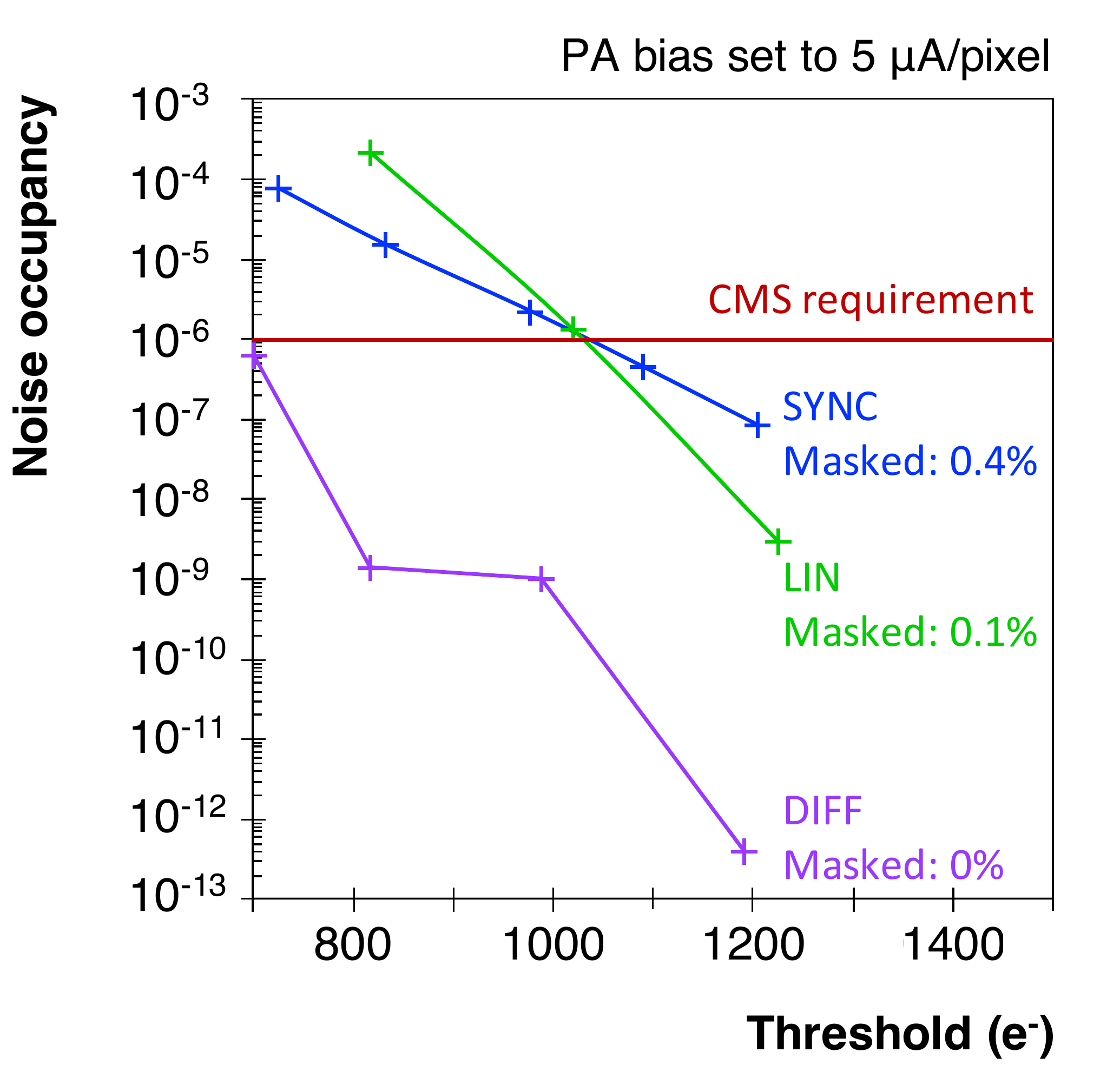}
        \caption{Influence of the threshold.}
        \label{fig:noise_thr}
    \end{subfigure}
    \hfill
    \begin{subfigure}{0.49\textwidth}
        \centering
        \includegraphics[width=\textwidth]{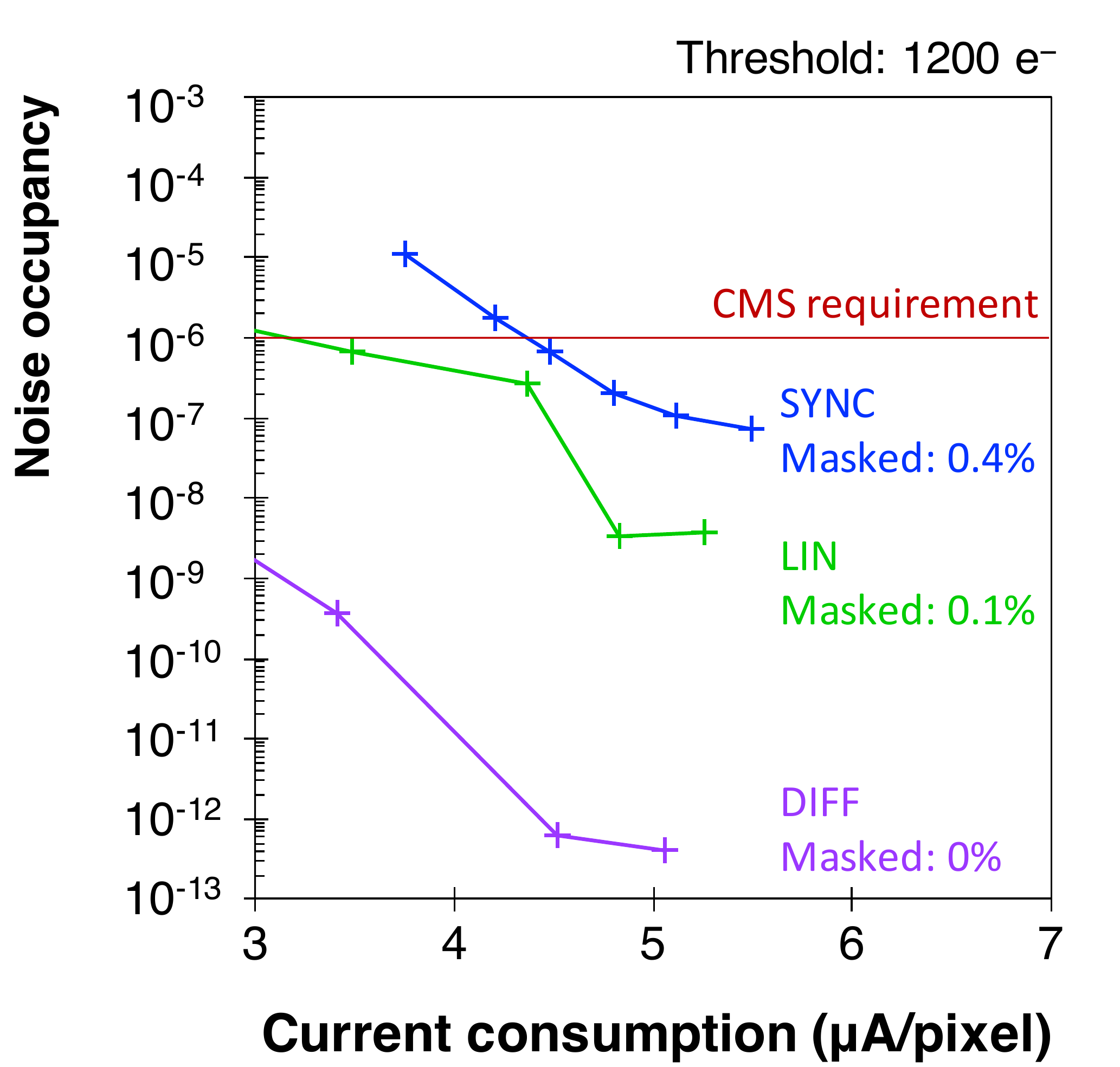}
        \caption{Influence of the \gls{pa} bias current \cite{luigi_vertex}.}
        \label{fig:noise_pabias}
    \end{subfigure}
    \caption{Noise occupancy measurement of the three \gls{afe}s implemented in the RD53A chip as a function of the threshold (a) and the \gls{pa} bias current (b). The \gls{sync} \gls{afe} is shown in blue, the \gls{lin} \gls{afe} in green and the \gls{diff} \gls{afe} in violet. The \gls{cms} requirement for the maximum noise occupancy is indicated in red. The number of masked noisy pixels is given for each \gls{afe}.}
    \label{fig:noise}
\end{figure}

The evaluation of the noise levels in the RD53A chip was done by sending triggers, without any charge injection, so that each recorded hit was induced by the noise. The average noise occupancy is then defined as the number of noise hits per pixel and per trigger. It was measured for the three \gls{afe}s, using a non-irradiated RD53A chip with a sensor with the highest capacitance, i.e.~rectangular pixels, operated at a temperature of about \SI{-10}{\celsius}.
Single noisy pixels can be disabled to reduce the rate of noise hits. 
However, the fraction of disabled pixels should not significantly increase the detection inefficiency.
A single pixel was considered noisy if its noise occupancy was above \num{e-4} based on the lowest occupancy expected in the \gls{it} detector from simulation (Figure~\ref{fig:occ}). Therefore, as a first step of the noise evaluation, every pixel with more than \num{100} hits in \num{e6} triggers was disabled at a threshold of \num{1200}~e${^{-}}$. 
As a second step, a new set of \num{e6} triggers was sent to each front-end to measure the noise occupancy of the non-masked pixels. In case of very low noise more triggers were sent. Results of noise occupancy measurements are presented in Figure~\ref{fig:noise}. The fraction of masked noisy pixels is indicated in the legend and the maximum noise occupancy of \num{e-6}, required by \gls{cms}, is indicated by the red line. The \gls{tot} was calibrated to \num{1100}~e$^{-}$/TOT\textsubscript{40}.

First, the influence of the threshold on the average noise occupancy was evaluated. The result is shown in Figure~\ref{fig:noise_thr}. The threshold was gradually decreased from \num{1200}~e${^{-}}$, keeping the same noisy pixels disabled. As expected, the average noise occupancy decreases with increasing threshold, regardless of the front-end design. The \gls{diff} front-end shows very good noise performance, with the average noise occupancy several orders of magnitude below the requirement, even for low thresholds. No hits were found in this front-end in \num{e6} triggers at higher thresholds, hence a higher number of triggers was sent to evaluate the average noise occupancy. The other two \gls{afe}s satisfy the noise requirement down to a threshold of \num{1000}~e${^{-}}$, which is consistent with the requirement on the detection threshold. Nonetheless, it can be noticed that the fraction of masked pixels is higher in the \gls{sync} \gls{afe}.

The influence of the \gls{pa} bias current on the noise was also studied. When this current increases, the transconductance of the input transistor is increased, which results in lower noise with a penalty of an increase in the analogue current consumption. The average noise occupancy was measured for different \gls{pa} bias currents and is presented in Figure~\ref{fig:noise_pabias} as a function of the measured analogue current consumption per pixel. All the other front-end settings that could contribute to the current consumption were kept constant during this measurement. 
As expected the noise in all three \gls{afe}s decreases when more current is provided.
The \gls{diff} \gls{afe} shows again a very good noise performance, with the average noise occupancy well below the requirement, even when operated with low \gls{pa} bias. The \gls{lin} and the \gls{sync} \gls{afe} need \SI{3.5}{\micro\ampere} and \SI{4.5}{\micro\ampere} per pixel, respectively, to reach the required noise level. All three \gls{afe}s can meet the \gls{cms} noise requirement if the \gls{pa} bias current is adjusted, hence this parameter is a handle to reduce the front-end noise at the price of an increase in the power consumption.
    
\section{Dead time and time-over-threshold calibration}
\label{sec:deadtime}

\begin{figure}[ht]
    \centering
    \includegraphics[width=0.55\textwidth]{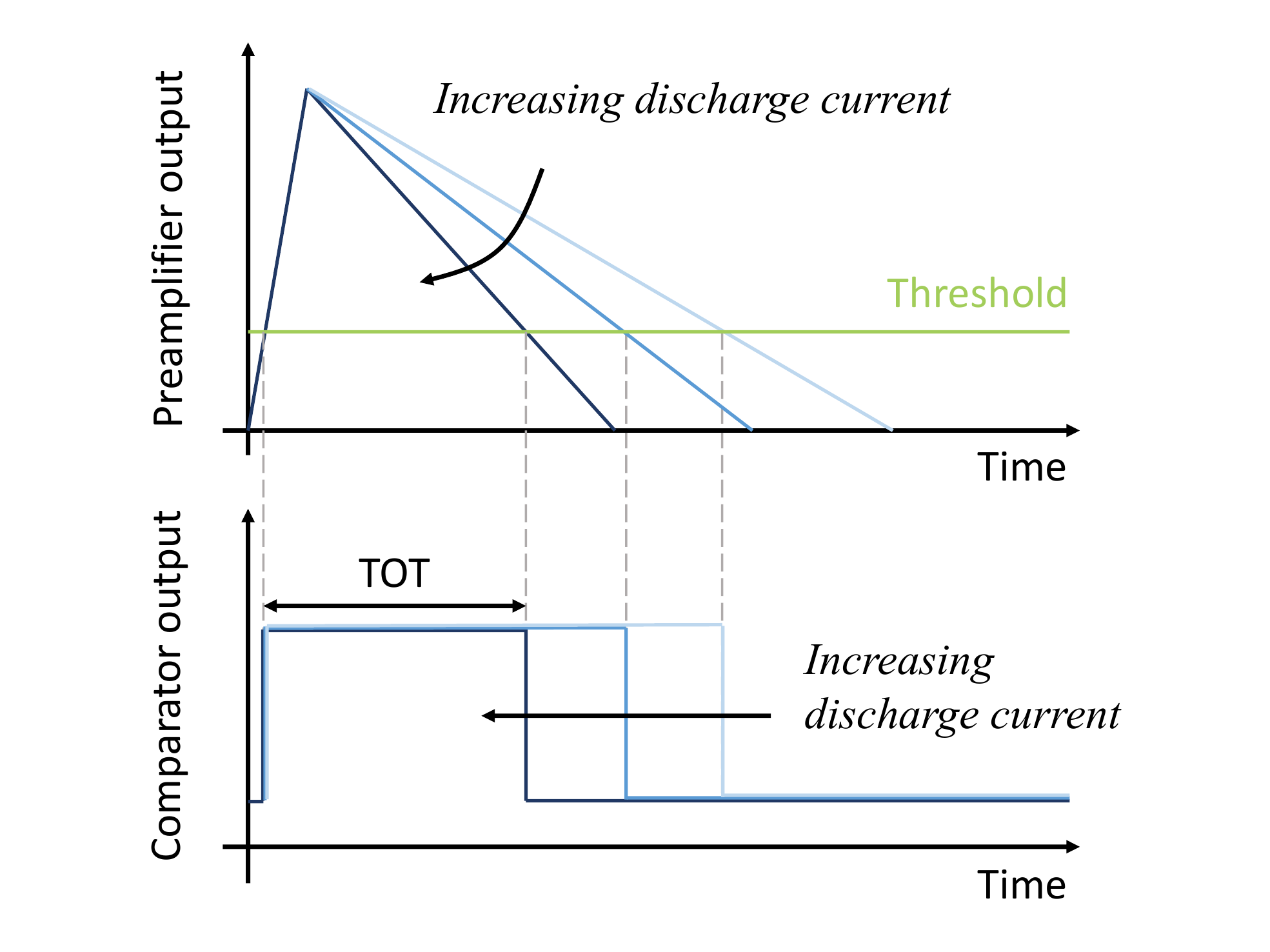}
    \caption{Sketch of the influence of the discharge current on the signal shape at the output of the \gls{pa} and on the corresponding \gls{tot}.}
    \label{fig:discharge-schetch}
\end{figure}

An important consideration for a highly efficient particle detector is the event loss due to the dead time, especially at high luminosity and high pileup.
As explained in Section~\ref{sec:requirements} the dead time caused by the \gls{afe} depends on the \gls{tot} calibration, and a charge resolution of \num{3000}~e${^{-}}$/TOT$_{40}$ is necessary to achieve the \SI{1}{\percent} dead time required for the innermost layer of the \gls{it} barrel.
The \gls{tot} response can be set by adjusting the discharge current of the \gls{pa}. When the \gls{pa} discharge current increases, the \gls{pa} output returns faster to the baseline and the corresponding \gls{tot} is smaller, as illustrated in Figure~\ref{fig:discharge-schetch}. Therefore, a faster \gls{pa} discharge leads to a reduced detector dead time. In the following the required charge resolution of \num{3000}~e${^{-}}$/TOT$_{40}$ is also referred to as the \emph{fast discharge}.

\begin{figure}[ht]
    \centering
    \begin{subfigure}{0.49\textwidth}
        \includegraphics[width=\textwidth]{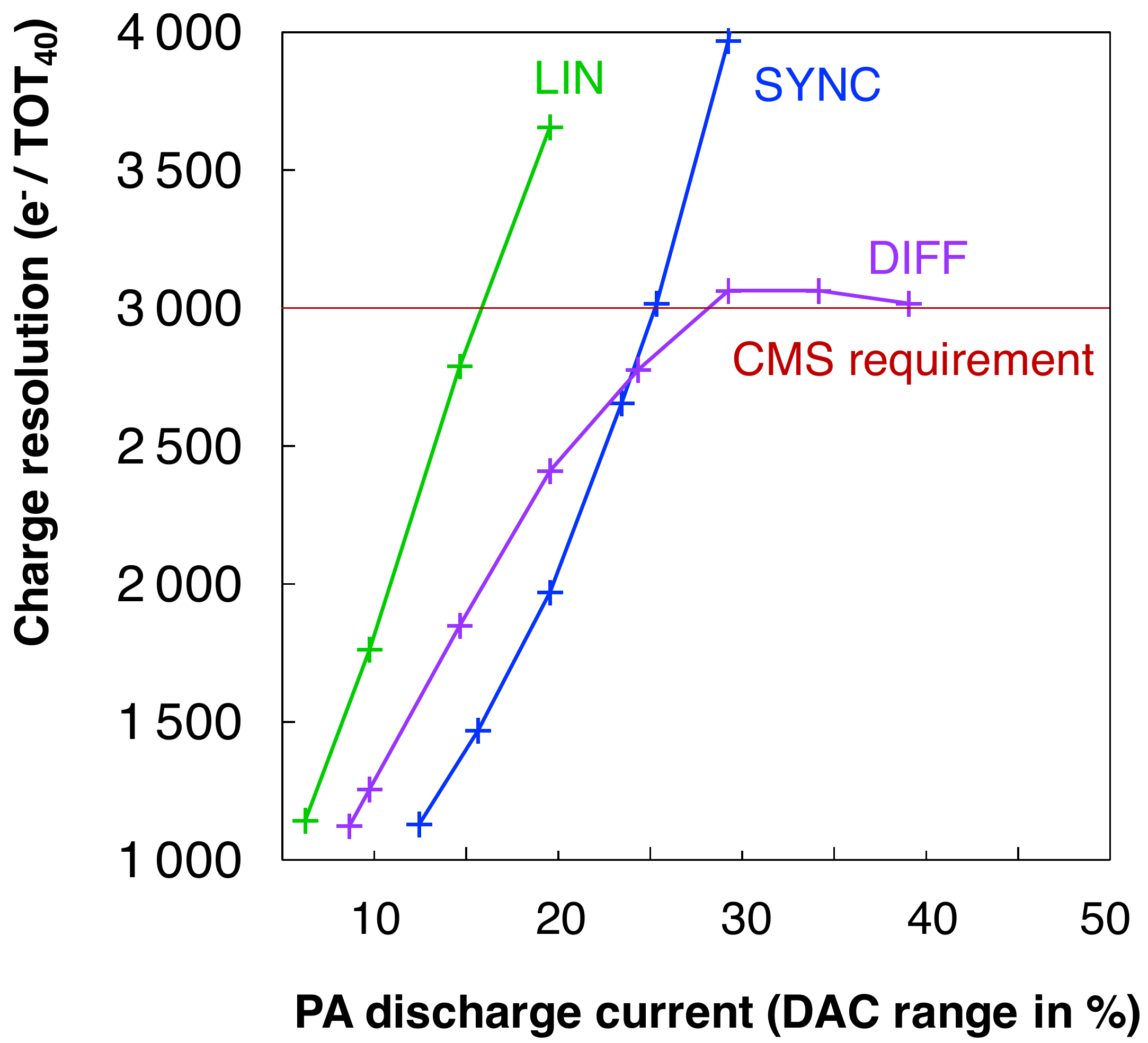}
        \caption{}
        \label{fig:dt}
    \end{subfigure}
    \hfill
    \begin{subfigure}{0.49\textwidth}
        \includegraphics[width=\textwidth]{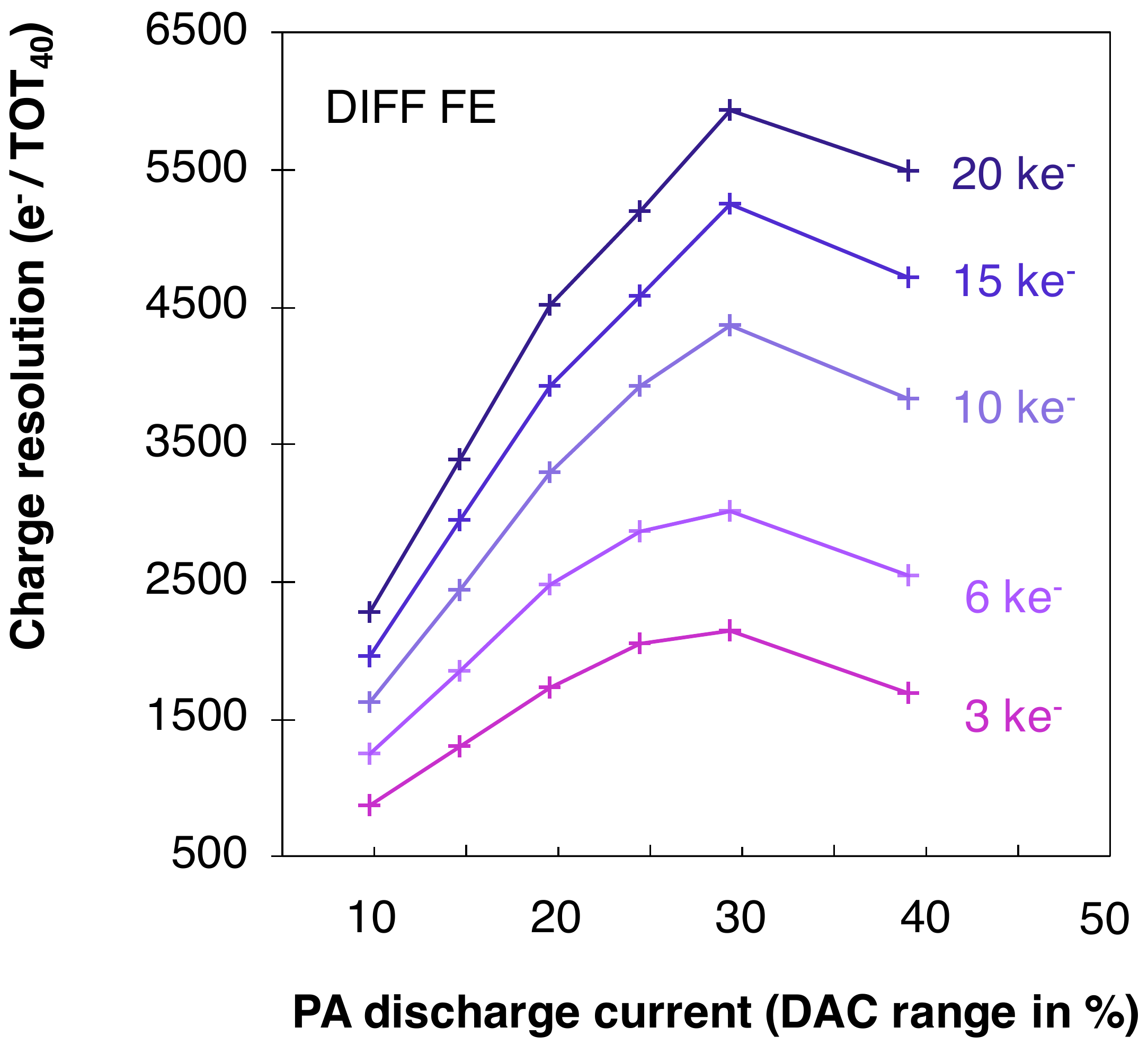}
        \caption{}
        \label{fig:dtdiff}
    \end{subfigure}  
    \caption{The charge resolution as a function of the \gls{pa} discharge current (a) measured with a constant charge injection of \num{6000}~e${^{-}}$ for the three RD53A \glspl{afe} and (b) measured for different input charges for the \gls{diff} \gls{afe} only.}
\end{figure}

The \gls{tot} charge resolution of the three \gls{afe}s was measured with a constant charge injection of \num{6000}~e${^{-}}$ for different \gls{pa} discharge currents. First, the charge resolution of all three \gls{afe}s was set to about \num{1100}~e${^{-}}$/TOT$_{40}$, as can be observed in Figure~\ref{fig:dt}. This resolution is not reached for the same current in different \gls{afe}s. 
In the next step, the \gls{pa} discharge current was increased to verify the front-end compliance with the dead time requirement. As expected, when the discharge current increases the \gls{pa} discharges faster and the charge resolution is coarser.
All three \gls{afe}s can reach the required charge resolution indicated by the red line. The \gls{sync} and \gls{lin} \glspl{afe} can also discharge faster, while the \gls{diff} \gls{afe} shows a saturation of the \gls{pa} discharge current \gls{dac} and would be operated at its limit to reach the dead time required for the inner layers.

A dedicated measurement was carried out on the \gls{diff} \gls{afe}, to better understand the observed saturation effect.
The charge resolution of the \gls{diff} \gls{afe} versus the discharge current was measured for different input charges, ranging from \num{3} to \num{20}~ke$^{-}$. The result, presented in Figure~\ref{fig:dtdiff}, confirms the saturation of the discharge current \gls{dac} in this \gls{afe}, occurring at \SI{30}{\percent} of the \gls{dac} range, regardless of the input charge. This implies a marginal operation of this particular \gls{afe} to reach the dead time requirement.

Increasing the discharge current reduces the dead time, as mentioned above, but it also reduces the \gls{afe} stability, and therefore it is likely to induce more noise. Hence the noise was re-evaluated for the fast discharge operation.
The noise was measured for two detection thresholds, \num{1000}~e${^{-}}$ and \num{1200}~e${^{-}}$, and two charge resolutions, \num{1100}~e${^{-}}$/TOT$_{40}$ and the required \num{3000}~e${^{-}}$/TOT$_{40}$.
The combination of these four parameters defined four measurement scenarios for which the average noise occupancy was measured. The measurement method was the same as in Section~\ref{sec:noise}. Pixels with more than \num{100} hits in \num{e6} triggers were declared noisy and masked, then the average noise occupancy of non-masked pixels was defined as the number of noise hits per pixel and per trigger, measured over \num{e6} events.

\begin{figure}[t]
    \centering
    \begin{subfigure}{0.49\textwidth}
        \centering
        \includegraphics[width=\textwidth]{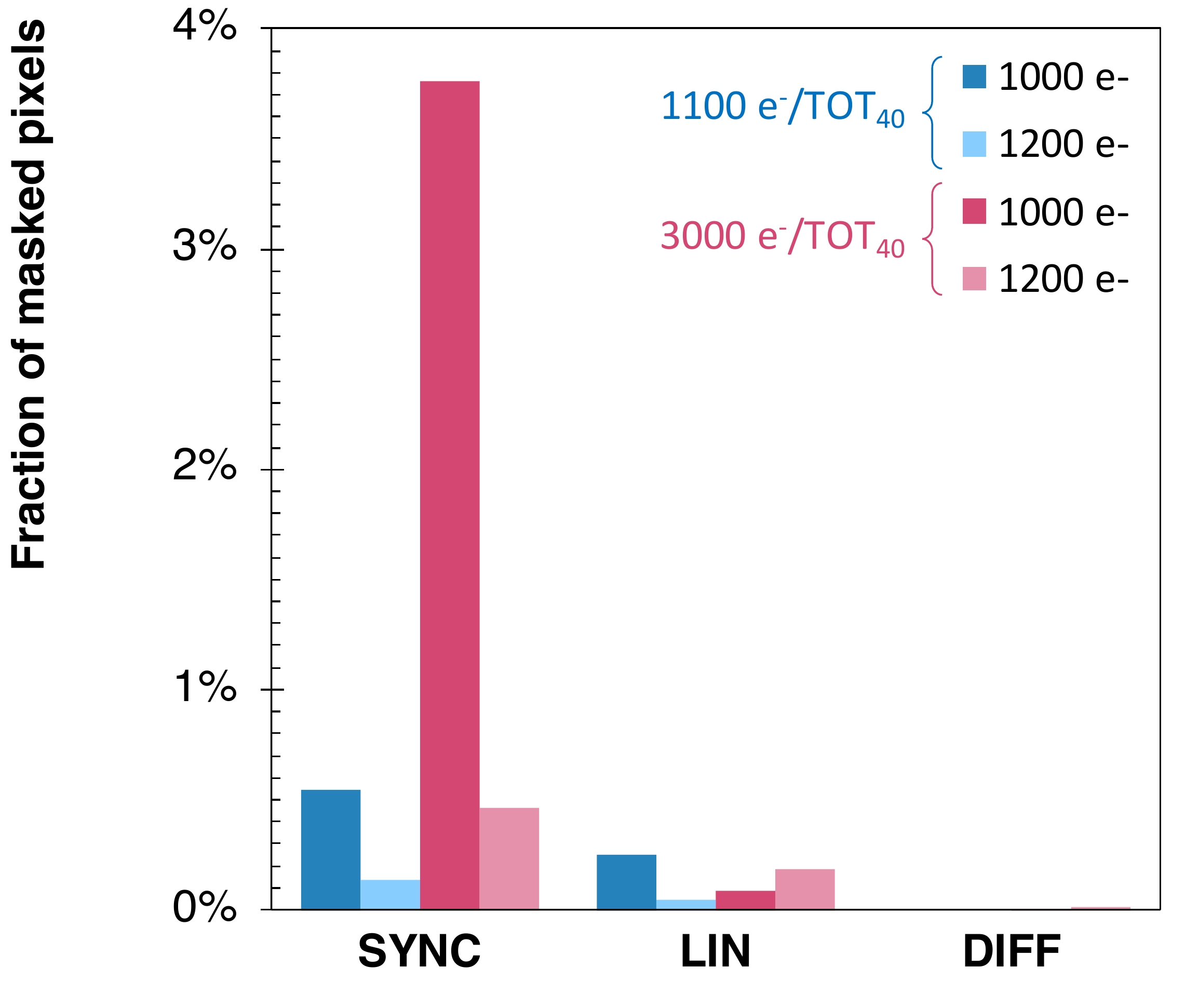}
        \caption{}
        \label{fig:dt-noisy-pixels}
    \end{subfigure}
    \hfill
    \begin{subfigure}{0.49\textwidth}
        \centering
        \includegraphics[width=\textwidth]{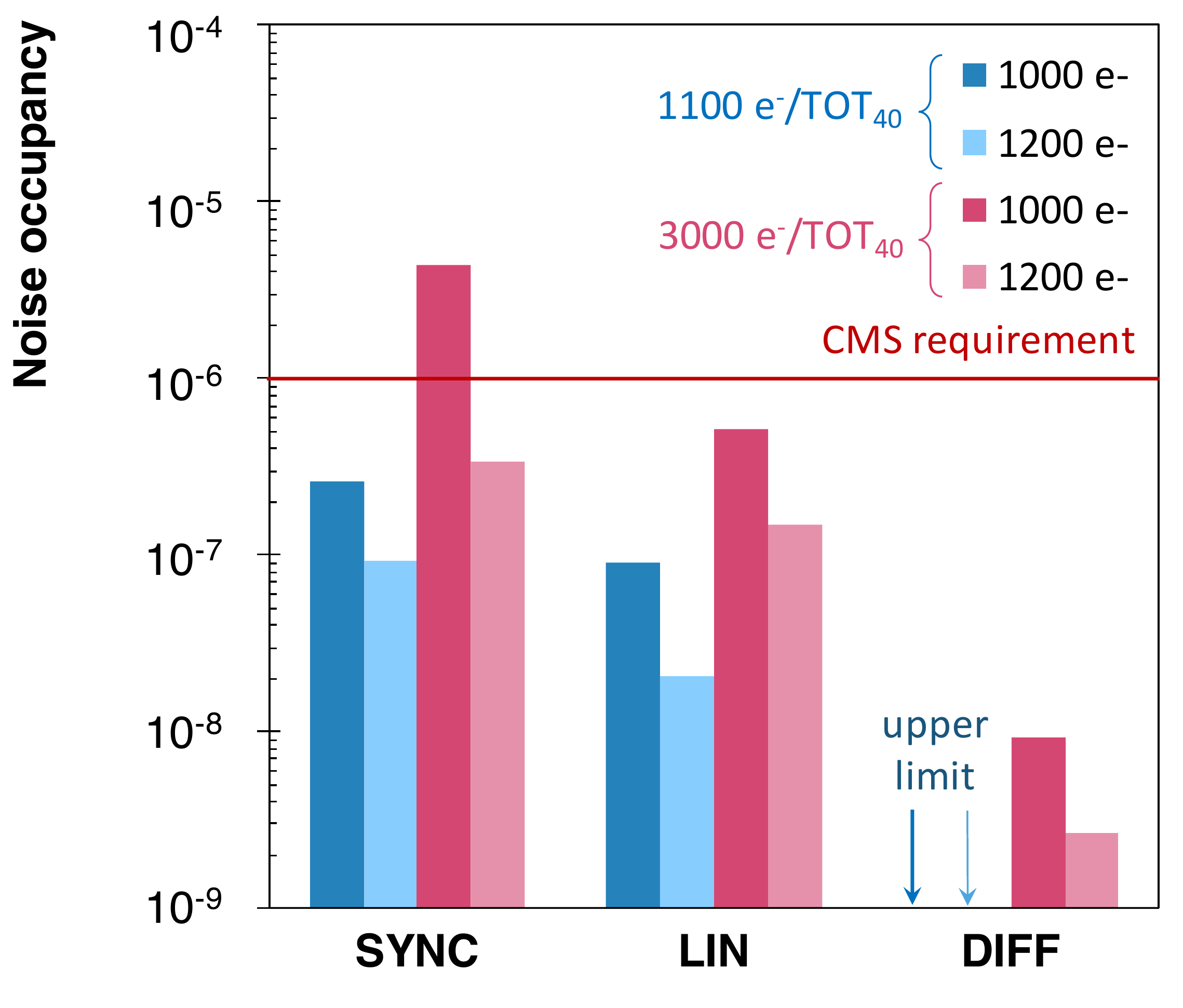}
        \caption{}
        \label{fig:dt-noise-occ}
    \end{subfigure}
    \caption{Fraction of masked noisy pixels~(a) and the average noise occupancy after masking~(b) of the three \gls{afe}s in the RD53A chip. The blue colours represent the charge calibration of \num{1100}~e${^{-}}$/TOT$_{40}$ and the red colours represent the charge calibration of \num{3000}~e${^{-}}$/TOT$_{40}$. The darker colours are used for the threshold of \SI{1000}{e^-} and the lighter colours are used for the threshold of \SI{1200}{e^-}.}
\end{figure}

The fraction of masked pixels is shown in Figure~\ref{fig:dt-noisy-pixels} and the average noise occupancy in Figure~\ref{fig:dt-noise-occ} for the four considered scenarios.
The average noise occupancy of all three \gls{afe}s is higher at fast discharge and is the highest at fast discharge and low threshold, as expected.
The \gls{diff} \gls{afe} demonstrates again excellent noise performance, with almost no noisy pixels and the average noise occupancy well below the requirement, even at fast discharge. At slow discharge the noise in this \gls{afe} was so low that only an upper limit was estimated. The \gls{lin} \gls{afe} has few noisy pixels and the average noise occupancy satisfies the requirement for any scenario. The \gls{sync} \gls{afe} appears to be the noisiest of the three, reaching almost \SI{3.8}{\percent} of noisy pixels when operated at fast discharge and low threshold. The higher noise in this \gls{afe}, significantly increasing with more aggressive chip settings, was considered a critical aspect for the operation in the innermost layer of the CMS Inner Tracker.

\section{Late-detected hits}
\label{sec:late}

\begin{figure}[t]
    \centering
    \captionsetup{justification=centering}
    \begin{minipage}{0.49\textwidth}
        \centering
        \includegraphics[width=0.89\textwidth]{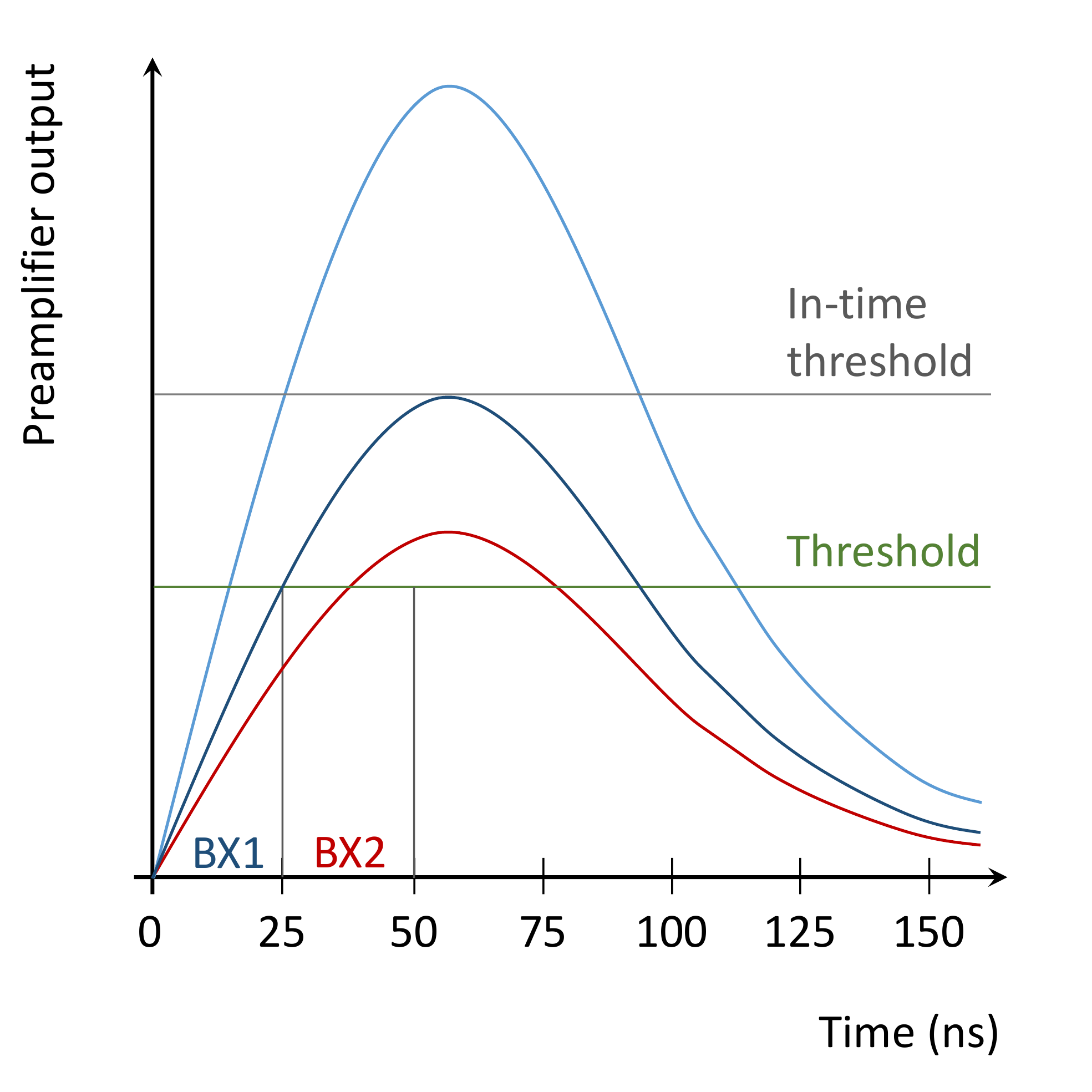}
        \caption{Illustration of the discriminator time response for different signal amplitudes.}
        \label{fig:late_hits_sketch}
    \end{minipage}
    \hfill
    \begin{minipage}{0.49\textwidth}
        \centering
        \includegraphics[width=0.89\textwidth]{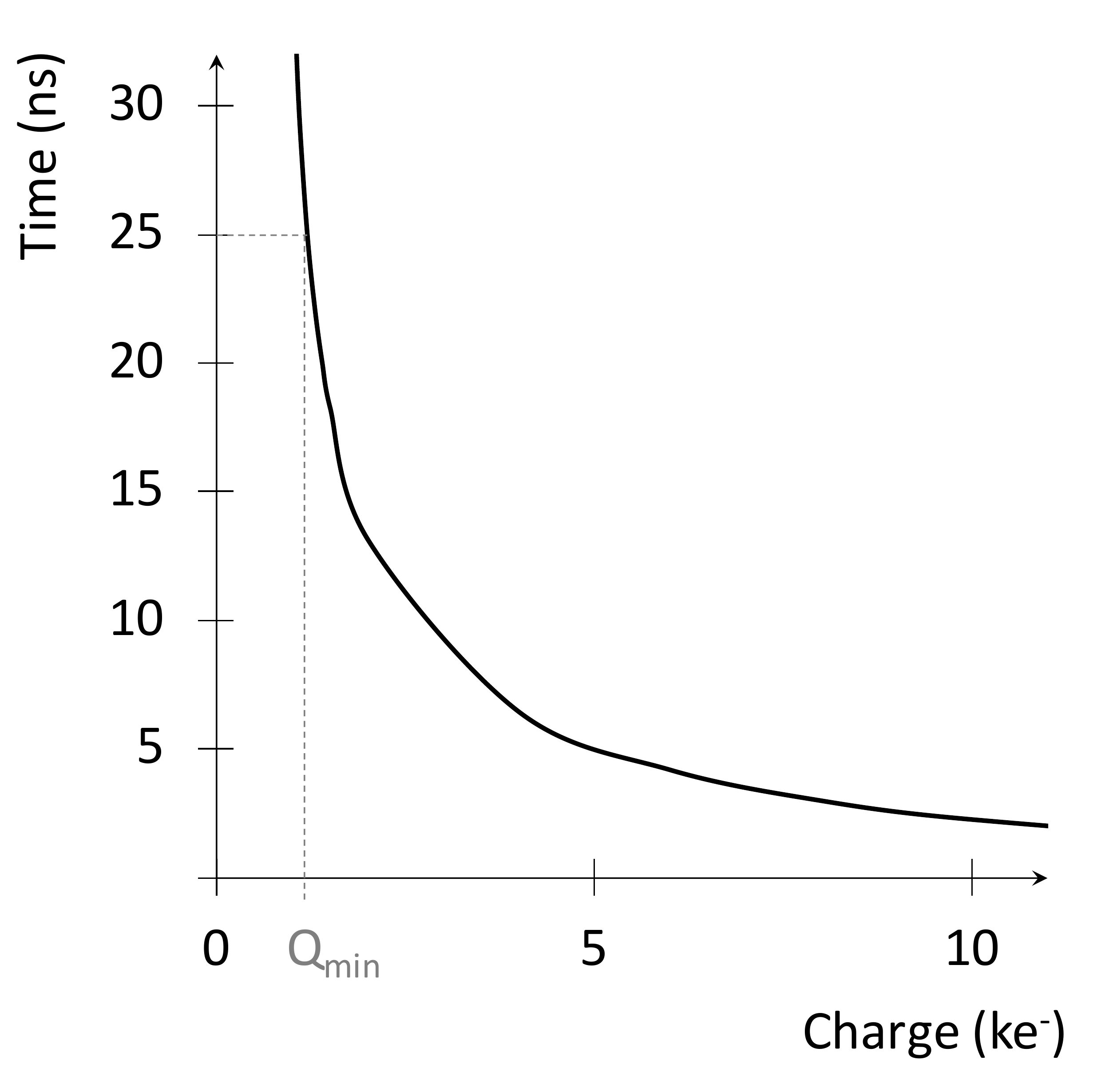}
        \caption{Simulated time walk curve of the linear front-end in the RD53A chip.}
        \label{fig:time_walk}
    \end{minipage}
\end{figure}

The time response of the pixel readout chip is important to assign detected hits to their corresponding \gls{lhc} \glspl{bx} and to limit the number of spurious hits from out-of-time pileup interactions.
The time response of the \gls{afe}, i.e.~the combination of the \gls{pa} rise time and the discriminator speed, is a function of the input charge. Pulses with the same peaking time but different amplitude pass the discriminator threshold at different times. High amplitude signals, depicted in light blue in Figure~\ref{fig:late_hits_sketch}, pass the threshold within one \gls{bx}, i.e.~within \SI{25}{\nano\second}. If the deposited charge is just above the threshold instead, the signal rises more slowly and is detected later by the discriminator. Such a hit, shown in red, might be assigned to the following \gls{bx}, and appears as a spurious hit in another event. The smallest charge ($\mathrm{Q_{min}}$) that can be detected within the correct \gls{bx} (dark blue signal) is equivalent to the so-called "in-time threshold", which is higher than the threshold of the discriminator. 
The time behaviour of an \gls{afe} is typically described by its time walk curve, i.e.~the response delay of the discriminator as a function of the input charge. An example of a simulated time walk curve of the \gls{lin} \gls{afe} is shown in Figure~\ref{fig:time_walk}. Given that the discriminator of the \gls{sync} \gls{afe} is locked to the clock, the following method is used to compare the timing of the three \glspl{afe}.


\subsection*{Front-end time response measurement}

\begin{figure}[ht]
    \centering
    \includegraphics[width=\textwidth]{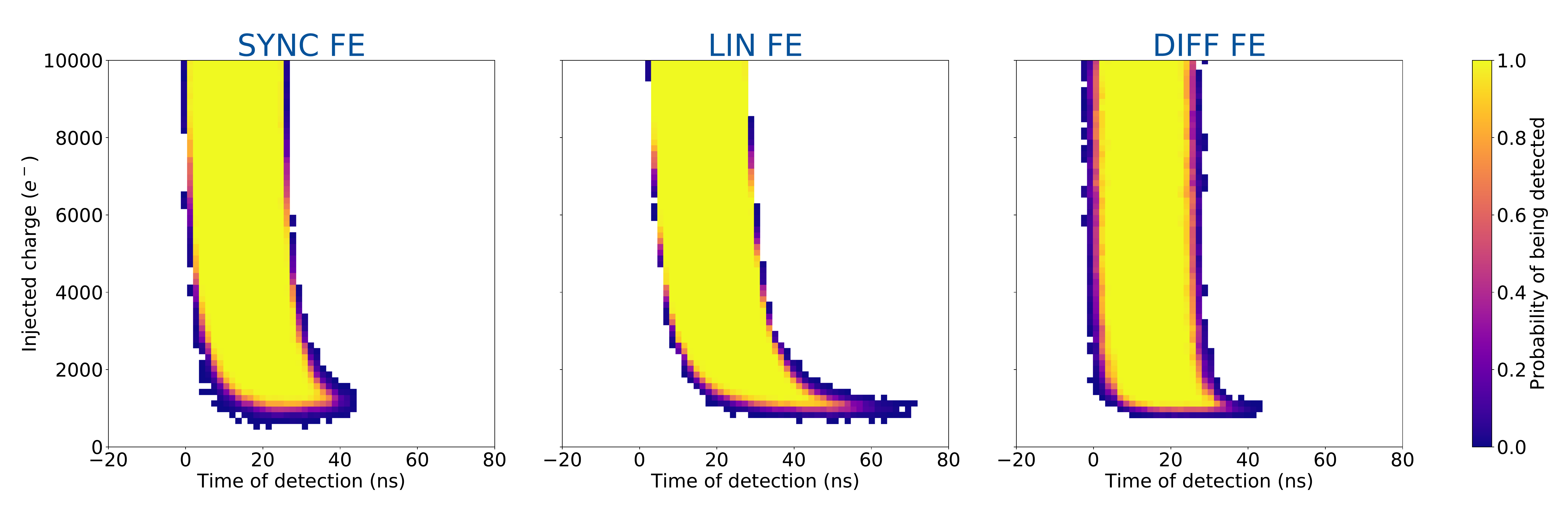}
    \caption{The measured time response of the \gls{sync} \gls{afe} (left), the \gls{lin} \gls{afe} (middle), and the \gls{diff} \gls{afe} (right) obtained with one RD53A chip.}
    \label{fig:3tor}
\end{figure}

The charge injection in the RD53A chip can be delayed with respect to the rising edge of the clock with a step size of \SI{1.5625}{\nano\second}~\citep{rd53a_manual}. The front-end time response was measured by injecting calibration pulses with different amplitudes and with different time delays. The detection threshold was set to \num{1000}~e${^{-}}$ and the full range of available charges up to \num{35}~ke${^{-}}$ was scanned, using a finer charge step for low charges where timing is critical. For every pixel the charge was injected \num{50} times for each time delay. Figure~\ref{fig:3tor} shows the two-dimensional plot of charge versus time for all three \glspl{afe}. 
The \hbox{$x$ axis} represents time in nanoseconds and $t=0$ indicates the time when the highest charge is detected. The \hbox{$y$ axis}, showing the injected charge in electrons, is limited to \num{10}~ke${^{-}}$ in this figure. The colour code indicates the detection probability for a given \gls{bx}, for each combination of charge and injection delay. The yellow zone corresponds to \SI{100}{\percent} detection efficiency, while in the white-coloured region no hit is detected. The left edge of the coloured region corresponds to the time walk curve.
In the upper part of the plot, the coloured region is a straight rectangle with a time width of \SI{25}{\nano\second}, which confirms that high charges are always detected within one \gls{bx}. Small charges instead are detected later, resulting in a tail in the detection region. This tail extends up to about \SI{40}{\nano\second} in the \gls{sync} and \gls{diff} \gls{afe}, indicating that these two \glspl{afe} have a comparable time response. The \gls{diff} \gls{afe} is able to correctly assign slightly smaller charges than the \gls{sync} \gls{afe}. 
The \gls{lin} \gls{afe} appears to be the slowest of the three, with the largest time walk of more than two \glspl{bx}.

\subsection*{Combination with time of arrival simulation}

\begin{figure}[ht]
    \centering
    \captionsetup{justification=centering}
    \begin{subfigure}{0.49\textwidth}
        \centering
        \includegraphics[width=\textwidth]{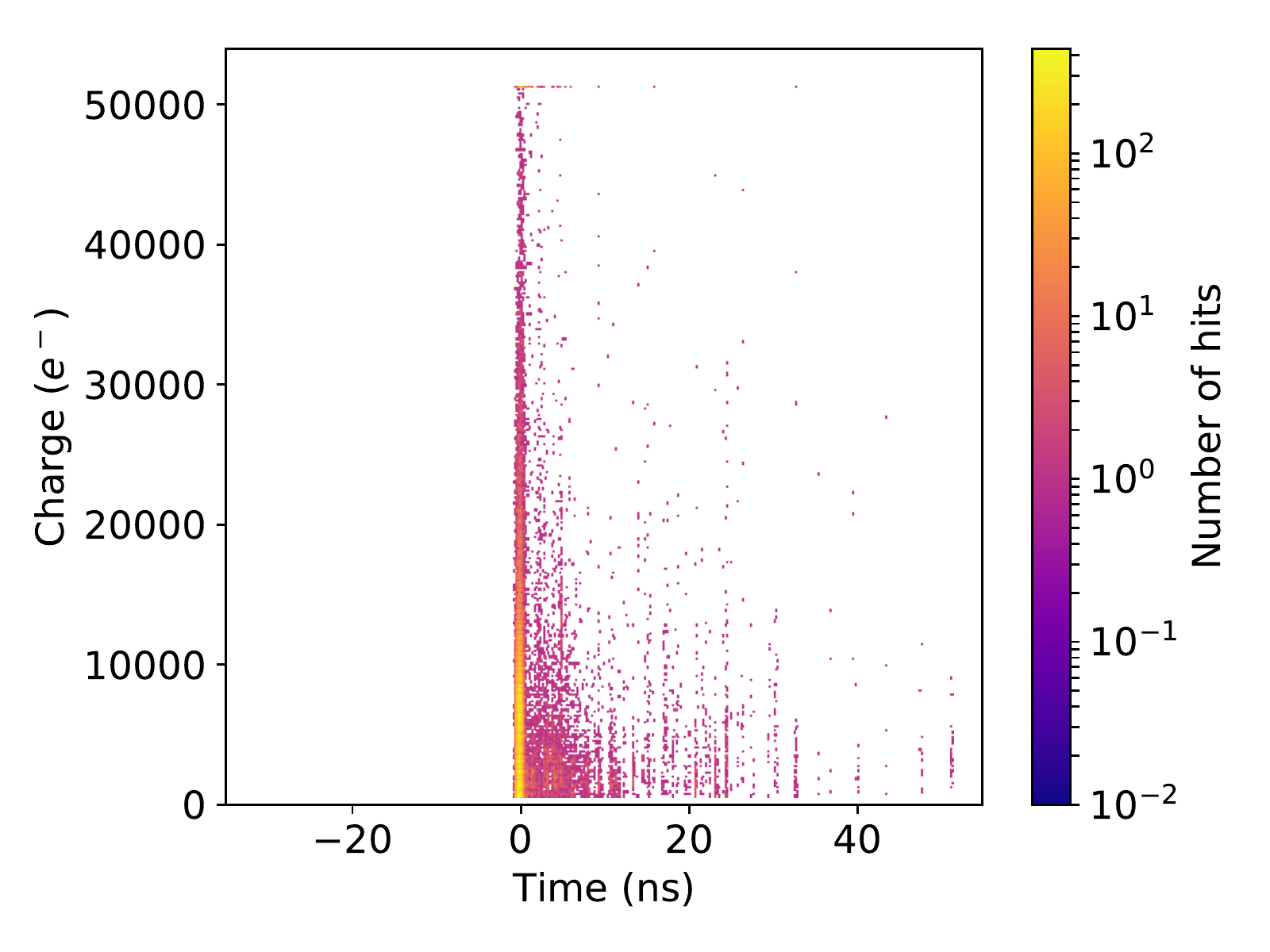}
        \caption{Time of arrival simulation.}
        \label{fig:mc}
    \end{subfigure}
    \hfill
    \begin{subfigure}{0.49\textwidth}
        \centering
        \includegraphics[width=\textwidth]{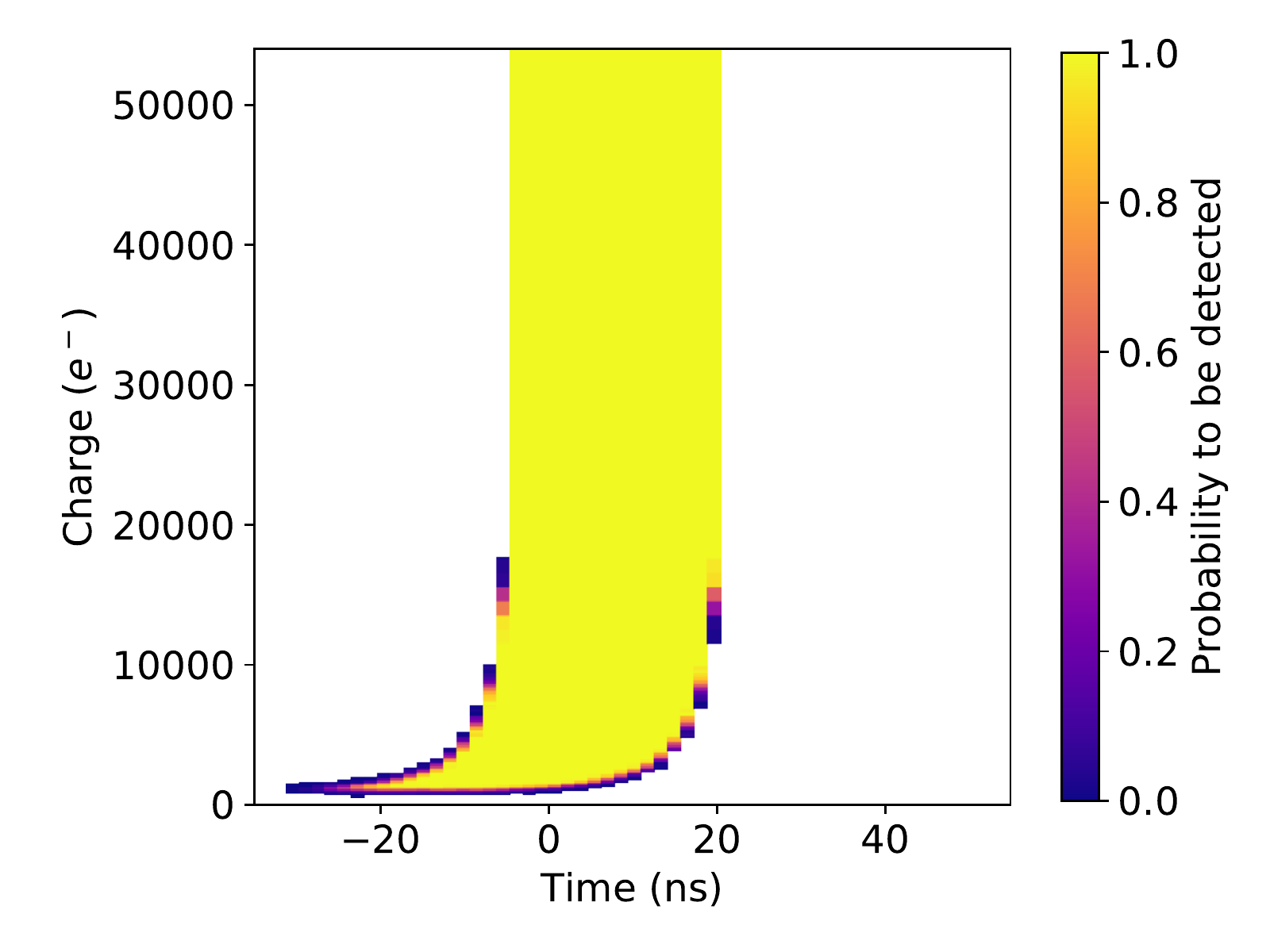}
        \caption{\gls{afe} acceptance region for a given \gls{bx}.}
        \label{fig:torproba}
    \end{subfigure}
    \begin{subfigure}{0.49\textwidth}
        \includegraphics[width=\textwidth]{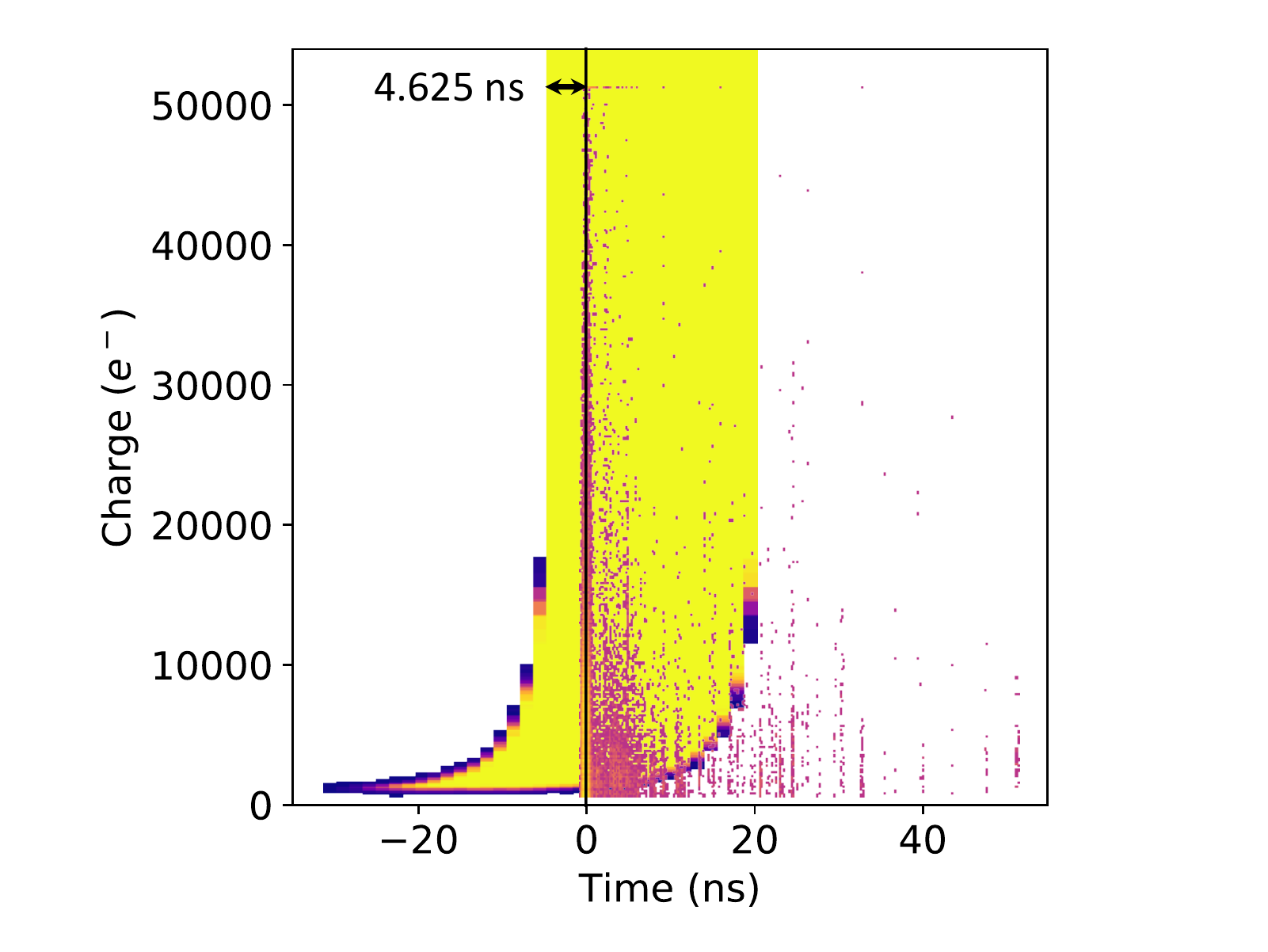}
        \caption{Overlay between measurement and simulation.}
        \label{fig:overlap}
    \end{subfigure}
    \hspace{2pt}
    \begin{subfigure}{0.49\textwidth}
        \centering
        \includegraphics[width=\textwidth]{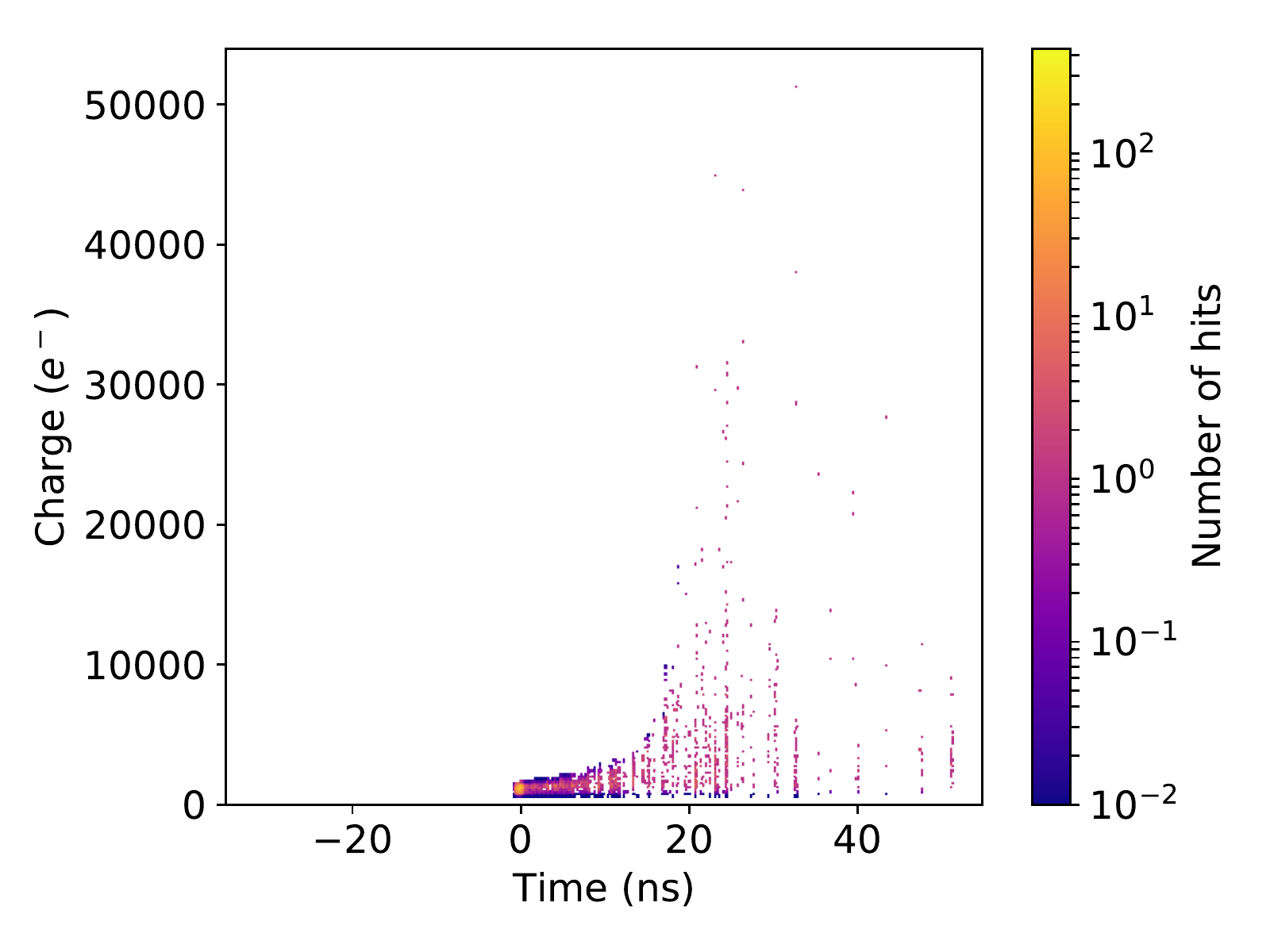}
        \caption{Late-detected hits.}
        \label{fig:beta}
    \end{subfigure}
    \caption{Different steps of the time response evaluation method.}
\end{figure}

A Monte Carlo simulation was performed within the standard \gls{cms} simulation and reconstruction software framework called CMSSW~\citep{cmssw} to evaluate the influence of the time response of each \gls{afe} on the detector performance and to estimate the resulting fraction of spurious hits. The time of arrival of particles was simulated for different locations in the \gls{it} detector, given that it depends on the position of the pixel module with respect to the interaction point. Sixteen different locations were studied: the centre ($z=0$) and edge module of each barrel layer, the innermost and outermost module of the first and last small disc, as well as of the first and last large disc (Figure~\ref{fig:itlayout}). For each location about \num{2000} minimum bias QCD events, without pileup and without a transverse momentum cut, were simulated. For the simulation of the beam spot a Gaussian distribution with a width ($\sigma$) in $z$ of \SI{4}{\centi\meter}, corresponding to a Gaussian width of \SI{130}{\pico\second} in time, was simulated.
The simulated pixel hits, corresponding to single pixels with deposited charge, were sorted by released charge, ranging from \num{600}~e${^{-}}$ \footnote{Given that charges smaller than \SI{600}{e^-} are not expected to be detected because of the threshold, the simulation started at this charge to avoid overloading the computing time.} to \num{50}~ke${^{-}}$, with a granularity of \num{150}~e${^{-}}$ and a time resolution of \SI{0.25}{\nano\second}. The simulated time of arrival versus charge distribution for the central module ($z=0$) of the innermost layer of the \gls{it} barrel is presented in Figure~\ref{fig:mc}.

Such a distribution was combined with the time response measurement introduced in the previous section.
The \hbox{$x$ axis} of the time response is reversed, obtaining the acceptance region, in time and charge, giving the probability of a charge to be detected in the correct \gls{bx}. This way, instead of showing when a hit is detected by the electronics, the figure indicates when a hit has to occur to be detected in a given \gls{bx}.
The \hbox{$y$ axis} has to be extended to match the charge range in the simulation. Assuming that the time response remains constant for very large signals, the yellow region with sharp edges is extended up to \num{50}~ke${^{-}}$.
For illustration, the time response of the \gls{lin} \gls{afe}, after such modifications, is presented in Figure~\ref{fig:torproba}.

When the acceptance region of the front-end is superimposed with the hit distribution from simulation, as indicated in Figure~\ref{fig:overlap}, the hits that are inside the yellow part of the acceptance region have a \SI{100}{\percent} probability to be assigned to the correct \gls{bx}. On the other hand, hits that are outside of the detection region have zero probability to be detected in time. Figure~\ref{fig:beta} shows the hits that will be assigned to a wrong \gls{bx}, obtained from the exclusion of the two overlaid plots. The integral of the exclusion plot divided by the total number of hits gives the fraction of late-detected hits in a given location of the future detector.

An important part of this method is the time alignment of the two overlapping plots. The origin of the time axis of both the measurement and the simulation have to be correctly aligned.
The $t=0$ of the simulation corresponds to the time when the two proton bunches overlap in the interaction region, corrected with the expected time-of-flight from the interaction point to the given module. The zero of the chip acceptance can be shifted to maximize the overlap, as it would be done in the detector by calibration. For this measurement, the peak of the simulation is placed three fully efficient bins from the left edge of the acceptance region, i.e.~\SI{4.625}{\nano\second}, as it is indicated in Figure~\ref{fig:overlap}. This estimate of about \SI{5}{\nano\second} was used to account for the imperfect time alignment in the detector due to the variations in the length of the electrical links, jitter, and other contributions, and also including some margin.

\subsection*{Fraction of late hits}

\begin{figure}[t]
    \centering
    \includegraphics[width=\textwidth]{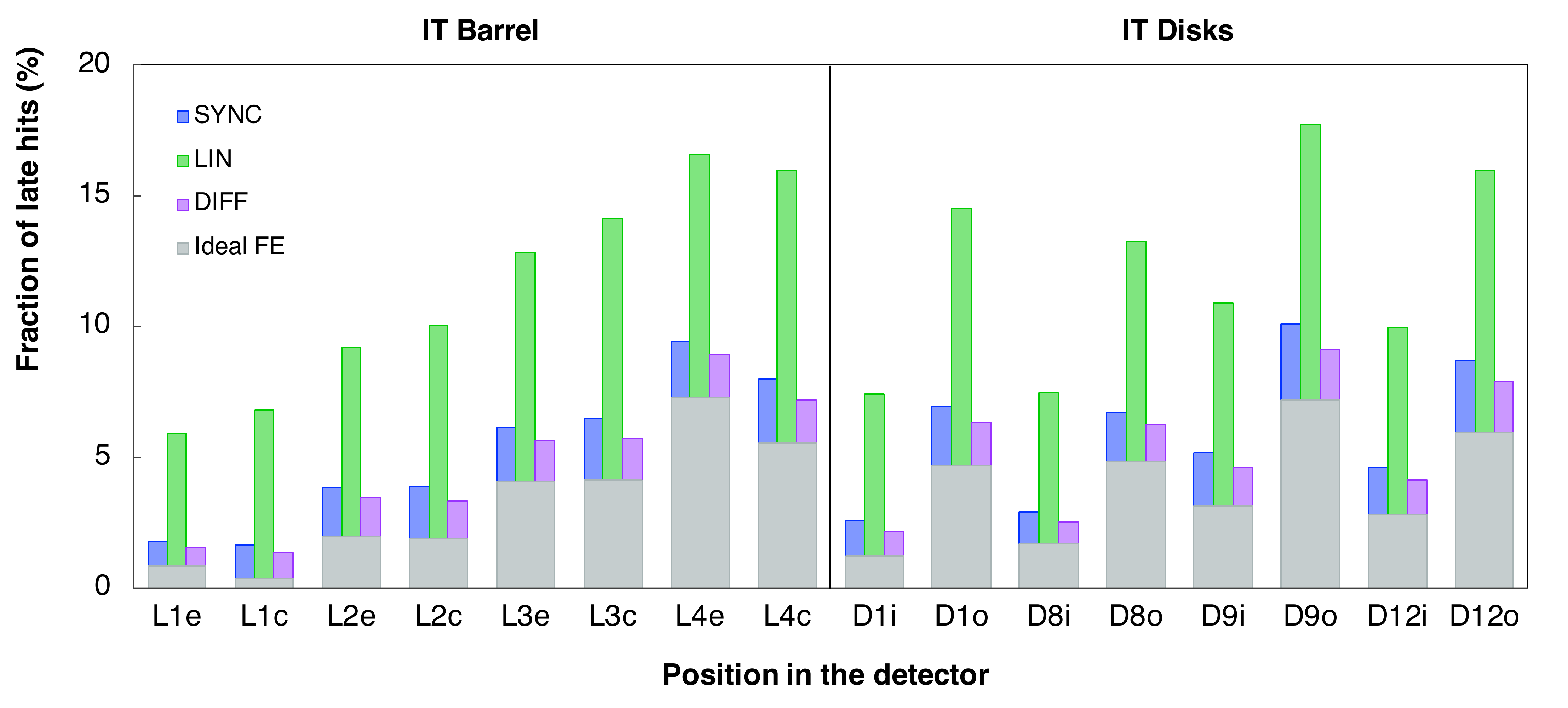}
    \caption{Fraction of hits detected late by the three RD53A \glspl{afe} for 16 pixel module positions. The \gls{it} barrel layers are numbered from the centre outwards L1 to L4 and for each layer the module at the edge is denoted "e" and the one in the centre is denoted "c". The \gls{it} discs are numbered D1 to D12 with increasing distance from the interaction point and for each disc the innermost ring is denoted "i" and the outermost one is denoted "o".}
    \label{fig:tor_late_hits}
\end{figure}

The method described above was used to evaluate the fraction of hits detected late by the three \gls{afe} designs. The result is shown in Figure~\ref{fig:tor_late_hits} for the selected detector locations.
The left half of the histogram corresponds to the \gls{it} barrel layers, numbered from the centre outwards L1 to L4. For each layer the study was done for two pixel modules, one at the edge (e) and the one in the centre (c) of the barrel. The fraction of late hits increases with the distance from the interaction point. 
The right half of the histogram is dedicated to the discs, numbered D1 to D12 with increasing distance from the interaction point. For each disc one module on the innermost (i) and one on the outermost (o) ring is presented. For any given disc the fraction of late detected hits is higher on the outer ring.

An ideal front-end with infinitely fast time response was also simulated 
and the fraction of hits detected late was estimated using the same method described above. Results are shown in grey in Figure~\ref{fig:tor_late_hits} overlaid to the estimates of the actual \acrlong{afe}s, because they represent the irreducible background. For the considered positions, this fraction is between \SI{0.38}{\percent} and \SI{7.26}{\percent}.
These are hits generated by particles whose travel time up to the sensor is more than \SI{25}{\nano\second} longer than the minimum, for which the detector is tuned. 
The \gls{sync} and \gls{diff} \gls{afe} have similar performance, causing few percent of misassigned hits on top of the background. The \gls{diff} is slightly faster.
The \gls{lin} \gls{afe} instead is significantly slower, causing up to additional \SI{11}{\percent} of late hits in the detector, on top of the irreducible \SI{7}{\percent} in the worst case.

\subsection*{LIN AFE slow time response mitigation}

Following the outcome of the previous measurement, a modification of the discriminator circuit was proposed by the design team to improve the time response of the \gls{lin} \gls{afe}. 
The discriminator is composed of two stages: a transconductance stage and a \acrfull{tia}. In the \gls{tia} two diode-connected transistors, initially introduced to minimise the static current consumption at the output of the discriminator, were forcing other transistors to operate in the deep sub-threshold regime, consequently making them slower. A significant improvement in time walk at the cost of a marginal increase in static current consumption was achieved by removing those two transistors. This led to a simpler \gls{tia} stage in the new design of the \gls{lin} \gls{afe}~\citep{new_lin}, for the next version of the chip, called RD53B~\citep{rd53b_manual}.

\begin{figure}[t]
    \centering
    \includegraphics[width=0.4\textwidth]{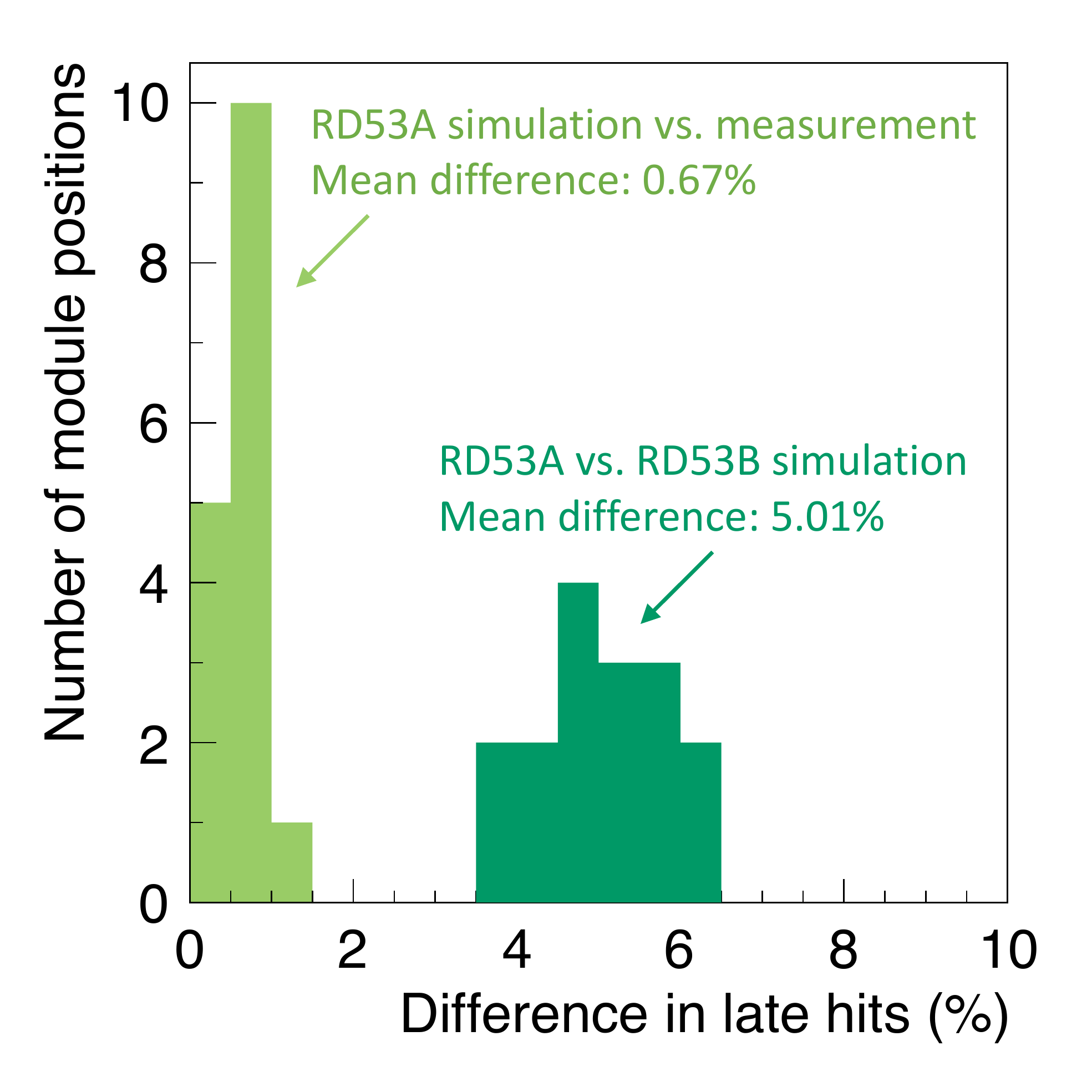}
    \caption{Difference between the fraction of hits detected late by the simulated and measured RD53A design of the \gls{lin} \gls{afe} (light green) and difference between the fraction of hits detected late by the simulated RD53A version of the \gls{lin} \gls{afe} and the improved RD53B version (dark green). }
    \label{fig:meas_vs_sim}
\end{figure}

Circuit simulations were used to extract the time walk curves of both the original and the improved \gls{lin} \gls{afe} designs. They were transformed into time response plots and combined with the time of arrival simulations to estimate the fraction of late hits for the simulated designs.
The simulated RD53A design was compared to the measurement and the difference in late hits is shown in Figure~\ref{fig:meas_vs_sim}. The simulated \gls{afe} gives a slightly higher number of late hits. Nevertheless, the simulation demonstrates a very good agreement with the measurement, the difference in late hits being below \SI{1}{\percent}. This confirms the validity of the simulation, which can therefore be used to predict the fraction of late hits in the improved design. The difference in late hits between the original design (RD53A) and the new one (RD53B) is also shown in Figure~\ref{fig:meas_vs_sim}. The new \gls{lin} \gls{afe} demonstrates on average \SI{5}{\percent} less misassigned hits. 
The improved design of the \gls{lin} \gls{afe} was also implemented in a test chip and verified before and after irradiation. The simulation and measurement results after an irradiation up to \SI{1}{\giga\rad} confirmed the improvement in time walk, which remains below \SI{20}{\nano\second}, whereas it increases to around \SI{30}{\nano\second} in the RD53A version~\citep{new_lin}.

\subsection*{Late-hit occupancy}

\begin{figure}[ht]
    \centering
    \includegraphics[width=\textwidth]{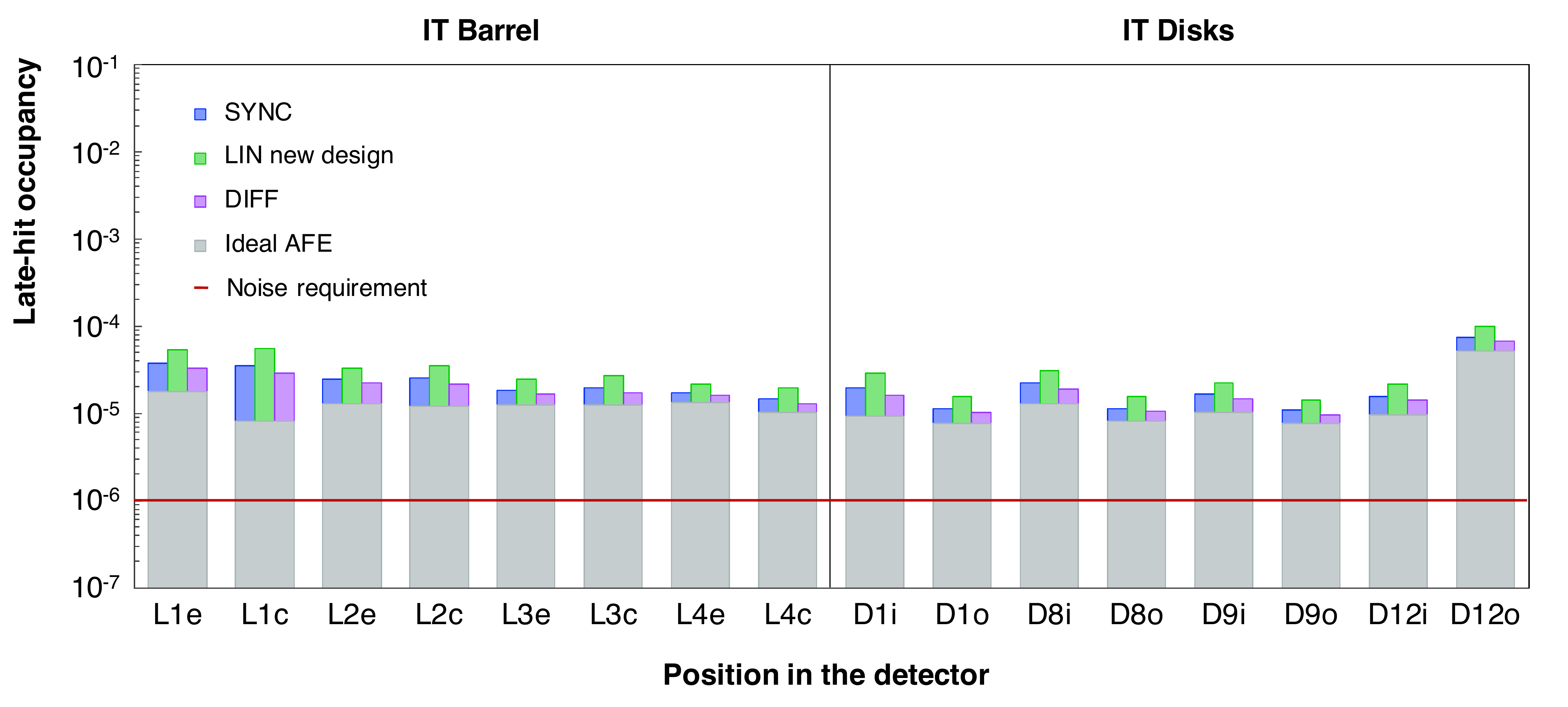}
    \caption{The occupancy due to hits detected late by the RD53A \glspl{afe} for different module positions in the detector. The \gls{it} barrel layers are numbered from the centre outwards L1 to L4 and for each layer the module at the edge is denoted "e" and the one in the centre is denoted "c". The \gls{it} discs are numbered D1 to D12 with increasing distance from the interaction point and for each disc the innermost ring is denoted "i" and the outermost one is denoted "o".}
    \label{fig:late-hit-occ}
\end{figure}

The fraction of late-detected hits was converted to the occupancy due to late hits, using the simulated hit occupancies extracted from Figure~\ref{fig:occ}. The result is shown in Figure~\ref{fig:late-hit-occ} for all the positions in the detector. The irreducible background of misassigned hits is almost uniform in the tracker and amounts to between \num{e-5} and \num{e-4}, regardless of the \gls{afe} design. The late-hit occupancy levels are at least one order of magnitude above the required noise level, indicated by the red line in the figure. Hence the spurious hit rate in the detector is dominated by the time response of the \gls{afe}, not by the noise.
Moreover, the performance of the improved design of the \gls{lin} \gls{afe} is comparable to the other two \glspl{afe}, although it remains slightly higher. 

    \section{Conclusions}

A new generation pixel readout chip is being designed for the upgrade of the \gls{cms} \acrlong{it} to cope with stringent requirements imposed by unprecedented radiation levels and hit rates. Three different analogue front-ends were designed by the RD53 Collaboration and implemented in a large scale demonstrator chip (RD53A). The three designs were characterized and the expected detector performance was evaluated against the requirements to choose the most suitable option for \gls{cms}.

The \acrlong{diff} \acrlong{afe} showed the best noise performance, with the noise occupancy several orders of magnitude below the requirement, as well as a very good time response. Nevertheless, this \acrlong{afe} showed a problematic threshold tuning after irradiation, at cold temperature (\SI{-10}{\celsius}) and with high leakage current. An improved design was proposed, expected to extend the operation with effective threshold tuning up to \SI{500}{\mega\rad}, according to the simulation results. 
A saturation in the preamplifier feedback requires operation at the limits of the \acrlong{tot} response in order to match the dead time requirement for the innermost layer of the \acrlong{it}. 

The \acrlong{sync} \acrlong{afe} features an automatic threshold tuning performed periodically by the auto-zeroing circuit and offers a very good timing performance. However, it appeared to be the noisiest of the three \acrlong{afe}s. The noise increased for lower thresholds and fast preamplifier return to baseline, becoming critical for the operation settings of Layer~\num{1}.

The \acrlong{lin} \acrlong{afe} satisfied all the requirements, but featured a slower time response.
However, an improved design was developed and is expected from simulation to reach a timing performance almost equivalent to the other two \acrlongpl{afe}. Since all the performance parameters of this \acrlong{afe} satisfy \gls{cms} requirements and the main drawback was addressed and mitigated, the \acrlong{lin} \acrlong{afe} was identified as the lowest-risk option for the future pixel detector. \gls{cms} selected the \acrlong{lin} \acrlong{afe} with an improved design for the integration into the next version of the RD53 pixel chip for \gls{cms} (the C-ROC). A prototype is expected to become available in 2021.

\section*{Acknowledgments}

The tracker groups gratefully acknowledge financial support from the following funding agencies: BMWFW and FWF (Austria); FNRS and FWO (Belgium); CERN; MSE and CSF (Croatia); Academy of Finland, MEC, and HIP (Finland); CEA and CNRS/IN2P3 (France); BMBF, DFG, and HGF (Germany); GSRT (Greece); NKFIA K124850, and Bolyai Fellowship of the Hungarian Academy of Sciences (Hungary); DAE and DST (India); IPM (Iran); INFN (Italy); PAEC (Pakistan); SEIDI, CPAN, PCTI and FEDER (Spain); Swiss Funding Agencies (Switzerland); MST (Taipei); STFC (United Kingdom); DOE and NSF (U.S.A.).

Individuals have received support from HFRI (Greece).
    \bibliography{ms}

\providecommand{\href}[2]{#2}\begingroup\raggedright\begin{thebibliography}{10}

\bibitem{hllhc}
G.~Apollinari, I.~Béjar~Alonso, O.~Brüning, M.~Lamont and L.~Rossi,
  \emph{{High-Luminosity Large Hadron Collider (HL-LHC)}},  Preliminary Design
  Report \href{https://cds.cern.ch/record/2116337}{CERN-2015-005}, 2015.

\bibitem{lhc}
L.~Evans and P.~Bryant, \emph{{LHC Machine}},
  \href{http://dx.doi.org/10.1088/1748-0221/3/08/S08001}{\emph{JINST}
  {\bfseries 3} (2008) S08001}.

\bibitem{cms_exp}
{\scshape CMS} Collaboration, S.~Chatrchyan et~al., \emph{The {CMS} experiment
  at the {CERN} {LHC}},
  \href{http://dx.doi.org/10.1088/1748-0221/3/08/S08004}{\emph{JINST}
  {\bfseries 3} (2008) S08004}.

\bibitem{hllhc_web}
{Official HL-LHC project website}.
  \url{https://project-hl-lhc-industry.web.cern.ch/content/project-schedule}.

\bibitem{cms_p2}
{CMS Collaboration}, \emph{{Technical Proposal for the Phase-II Upgrade of the
  CMS Detector}},  Tech. Rep.
  \href{https://cds.cern.ch/record/2020886?ln=fr}{CERN-LHCC-2015-010,
  LHCC-P-008, CMS-TDR-15-02}, 2015.

\bibitem{p2_tdr}
{CMS Collaboration}, \emph{{The Phase-2 Upgrade of the CMS Tracker}},  Tech.
  Rep. \href{https://cds.cern.ch/record/2272264}{CERN-LHCC-2017-009,
  CMS-TDR-014}, 2017.

\bibitem{bril_cdr}
{CMS Collaboration}, \emph{{The Phase-2 Upgrade of the CMS Beam Radiation,
  Instrumentation, and Luminosity Detectors: Conceptual Design}},  Tech. Rep.
  \href{https://cds.cern.ch/record/2706512}{CMS-NOTE-2019-008,
  CERN-CMS-NOTE-2019-008}, 2020.

\bibitem{sp_vertex19}
{D. Koukola on behalf of the CMS Collaboration}, \emph{{Serial Powering for the
  Tracker Phase-2 Upgrade}},
  \href{http://dx.doi.org/10.22323/1.373.0039}{\emph{PoS} (Vertex2019) 039}.

\bibitem{L1_trigger}
{CMS Collaboration}, \emph{{The Phase-2 Upgrade of the CMS Level-1 Trigger}},
  Tech. Rep. \href{http://cds.cern.ch/record/2714892}{CERN-LHCC-2020-004,
  CMS-TDR-021}, 2020.

\bibitem{bane_module}
{B. Ristic on behalf of the CMS Tracker Group}, \emph{{Prototype Module
  Construction for the High Luminosity Upgrade of the CMS Pixel Detector}},
  \href{http://dx.doi.org/10.22323/1.373.0058}{\emph{PoS} (Vertex2019) 058}.

\bibitem{tsmc_web}
{Taiwan Semiconductor Manufacturing Company (TSMC) website}.
  \url{https://www.tsmc.com/}.

\bibitem{rd53}
{\scshape RD53} Collaboration, J.~Christiansen and M.~Garcia-Sciveres,
  \emph{{RD Collaboration Proposal: Development of pixel readout integrated
  circuits for extreme rate and radiation}},  Scientific Committee Paper
  \href{http://cds.cern.ch/record/1553467}{CERN-LHCC-2013-008, LHCC-P-006},
  2013.

\bibitem{rd53a_manual}
{\scshape RD53} Collaboration, M.~Garcia-Sciveres, \emph{{The RD53A Integrated
  Circuit}},  Tech. Rep.
  \href{https://cds.cern.ch/record/2287593}{CERN-RD53-PUB-17-001}, 2017.

\bibitem{rd53a_specs}
{\scshape RD53} Collaboration, M.~Garcia-Sciveres, \emph{{RD53A Integrated
  Circuit Specifications}},  Tech. Rep.
  \href{https://cds.cern.ch/record/2113263}{CERN-RD53-PUB-15-001}, 2015.

\bibitem{rd53a_cdr}
K.~Moustakas, P.~Rymaszewski, T.~Hemperek, H.~Krüger, M.~Vogt, T.~Wang and
  N.~Wermes, \emph{{A Clock and Data Recovery Circuit for the ATLAS/CMS HL-LHC
  Pixel Front End Chip in 65 nm CMOS Technology}},
  \href{http://dx.doi.org/10.22323/1.370.0046}{\emph{PoS} (TWEPP2019) 046}.

\bibitem{rd53a_transmitter}
T.~Wang, T.~Hemperek, H.~Krüger, K.~Moustakas, P.~Rymaszewski and M.~Vogt,
  \emph{{A high speed transmitter circuit for the ATLAS/CMS HL-LHC pixel
  readout chip}}, \href{http://dx.doi.org/10.22323/1.343.0098}{\emph{PoS}
  (TWEPP2018) 098}.

\bibitem{shldo}
M.~{Karagounis}, D.~{Arutinov}, M.~{Barbero}, F.~{Huegging}, H.~{Krueger} and
  N.~{Wermes}, \emph{{An integrated Shunt-LDO regulator for serial powered
  systems}},
  \href{http://dx.doi.org/10.1109/ESSCIRC.2009.5325974}{\emph{Proceedings of
  ESSCIRC} (2009) 276}.

\bibitem{rd53b_manual}
{\scshape RD53} Collaboration, M.~Garcia-Sciveres, F.~Loddo and
  J.~Christiansen, \emph{{RD53B Manual}},  Tech. Rep.
  \href{https://cds.cern.ch/record/2665301}{CERN-RD53-PUB-19-002}, 2019.

\bibitem{krum}
F.~Krummenacher, \emph{{Pixel detectors with local intelligence: an IC designer
  point of view}},
  \href{http://dx.doi.org/https://doi.org/10.1016/0168-9002(91)90152-G}{\emph{NIM
  A} {\bfseries 305} (1991) 527}.

\bibitem{sync}
E.~Monteil, L.~Pacher, A.~Paternò, N.~Demaria, A.~Rivetti, M.~Da~Rocha~Rolo,
  F.~Rotondo et~al., \emph{{A synchronous analog very front-end in 65 nm CMOS
  with local fast ToT encoding for pixel detectors at HL-LHC}},
  \href{http://dx.doi.org/10.1088/1748-0221/12/03/C03066}{\emph{JINST}
  {\bfseries 12} (2017) C03066}.

\bibitem{lin}
L.~Gaioni, F.~De~Canio, M.~Manghisoni, L.~Ratti, V.~Re and G.~Traversi,
  \emph{{65 nm CMOS analog front-end for pixel detectors at the HL-LHC}},
  \href{http://dx.doi.org/10.1088/1748-0221/11/02/C02049}{\emph{JINST}
  {\bfseries 11} (2016) C02049}.

\bibitem{timon_hiroshima}
T.~Heim, \emph{{Test results from the RD53A pixel readout chip and design
  status of its successor}},  in \emph{Proceedings of The 12th International
  Hiroshima Symposium on the Development and Application of Semiconductor
  Tracking Detectors}, 2019.

\bibitem{luigi_vertex}
L.~Gaioni, \emph{{RD53 analog front-end processors for the ATLAS and CMS
  experiments at the High-Luminosity LHC}},
  \href{http://dx.doi.org/10.22323/1.373.0021}{\emph{PoS} (Vertex2019) 021}.

\bibitem{aleksandra_vci}
A.~Dimitrievska and A.~Stiller, \emph{{RD53A: A large-scale prototype chip for
  the phase II upgrade in the serially powered HL-LHC pixel detectors}},
  \href{http://dx.doi.org/10.1016/j.nima.2019.04.045}{\emph{NIM A} {\bfseries
  958} (2020) 162091}.

\bibitem{ennio_twepp}
E.~Monteil, \emph{{RD53A: a large scale prototype for HL-LHC silicon pixel
  detector phase 2 upgrades}},
  \href{http://dx.doi.org/10.22323/1.343.0157}{\emph{PoS} (TWEPP2018) 157}.

\bibitem{luigi_nima}
L.~Gaioni, \emph{{Test results and prospects for RD53A, a large scale 65 nm
  CMOS chip for pixel readout at the HL-LHC}},
  \href{http://dx.doi.org/10.1016/j.nima.2018.11.107}{\emph{NIM A} {\bfseries
  936} (2019) 282}.

\bibitem{bdaq53}
M.~Daas, Y.~Dieter, J.~Dingfelder, M.~Frohne, G.~Giakoustidis, T.~Hemperek,
  F.~Hinterkeuser et~al., \emph{{BDAQ53, a versatile pixel detector readout and
  test system for the ATLAS and CMS HL-LHC upgrades}},
  \href{http://dx.doi.org/https://doi.org/10.1016/j.nima.2020.164721}{\emph{NIM
  A} {\bfseries 986} (2021) 164721}.

\bibitem{p1_tdr}
{CMS Collaboration}, \emph{{CMS Technical Design Report for the Pixel Detector
  Upgrade}},  Tech. Rep.
  \href{http://cds.cern.ch/record/1481838}{CERN-LHCC-2012-016, CMS-TDR-11},
  2012.

\bibitem{kit}
V.~Cindro, \emph{{World Irradiation Facilities for Silicon Detectors}},
  \href{http://dx.doi.org/10.22323/1.227.0026}{\emph{PoS} (Vertex2014) 026}.

\bibitem{cmssw}
D.~J. Lange, \emph{The {CMS} reconstruction software},
  \href{http://dx.doi.org/10.1088/1742-6596/331/3/032020}{\emph{Journal of
  Physics: Conference Series} {\bfseries 331} (2011) 032020}.

\bibitem{new_lin}
{\scshape RD53} Collaboration, L.~Gaioni and F.~Loddo, \emph{{CMS analog
  front-end: simulations and measurements}},  Tech. Rep.
  \href{https://cds.cern.ch/record/2746420}{CERN-RD53-PUB-20-002}, 2020.

\end{thebibliography}\endgroup
    
\newcommand{\cmsAuthorMark}[1]
{\hbox{\textsuperscript{\normalfont#1}}}

\setlength{\emergencystretch}{3em}  
\providecommand{\tightlist}{%
  \setlength{\itemsep}{0pt}\setlength{\parskip}{0pt}}
\setcounter{secnumdepth}{0}
\ifx\paragraph\undefined\else
\let\oldparagraph\paragraph
\renewcommand{\paragraph}[1]{\oldparagraph{#1}\mbox{}}
\fi
\ifx\subparagraph\undefined\else
\let\oldsubparagraph\subparagraph
\renewcommand{\subparagraph}[1]{\oldsubparagraph{#1}\mbox{}}
\fi

\newpage
\section*{\Large The Tracker Group of the CMS Collaboration}

\addcontentsline{toc}{section}{The Tracker Group of the CMS Collaboration}
\setlength{\parindent}{0pt}
\setlength{\parskip}{8pt plus 2pt minus 1pt}

\textcolor{black}{\textbf{Institut~f\"{u}r~Hochenergiephysik, Wien, Austria}\newline 
W.~Adam, T.~Bergauer, D.~Bl\"{o}ch, M.~Dragicevic, R.~Fr\"{u}hwirth\cmsAuthorMark{1}, V.~Hinger, H.~Steininger}

\textcolor{black}{\textbf{Universiteit~Antwerpen, Antwerpen, Belgium}\newline
W.~Beaumont, D.~Di~Croce, X.~Janssen, T.~Kello, A.~Lelek,  P.~Van~Mechelen, S.~Van~Putte, N.~Van~Remortel
% ...
}

\textcolor{black}{\textbf{Texas~A\&M~University, College~Station, USA}\newline
R.~Eusebi}

\textcolor{black}{\textbf{Vanderbilt~University, Nashville, USA}\newline
P.~D'Angelo, W.~Johns}

\dag: Deceased\\
1: Also at Vienna University of Technology, Vienna, Austria \\
2: Also at Institute for Theoretical and Experimental Physics, Moscow, Russia \\
3: Also at Universit\'{e} de Haute-Alsace, Mulhouse, France \\
4: Also at Horia Hulubei National Institute of Physics and Nuclear Engineering~(IFIN-HH), Bucharest, Romania \\
5: Also at University of Cagliari, Cagliari, Italy \\
6: Also at Institut Polytechnique de Grenoble, Grenoble, France \\
7: Also at Universit\'{e}~de~Strasbourg, CNRS, IPHC~UMR~7178, Strasbourg, France \\
8: Also at Universidad de Castilla-La-Mancha, Ciudad Real, Spain 
\\
9: Also at University of Patras, Patras, Greece \\
10: Also at \'{E}cole Polytechnique F\'{e}d\'{e}rale de Lausanne, Lausanne, Switzerland \\
11: Also at National Technical University of Athens, Athens, Greece \\
12: Also at Universit\"{a}t~Z\"{u}rich,~Zurich,~Switzerland \\
13: Now at Bethel University, St. Paul, Minnesota, USA \\
14: Now at Karamanoglu Mehmetbey University, Karaman, Turkey \\
15: Now at Florida State University, Tallahassee, USA \\
16: Also at Nanjing Normal University, Nanjing, China
\end{document}